\documentclass[traditabstract]{aa}
\usepackage{amsmath,amssymb,textcomp}
\usepackage[tight]{subfigure}
\usepackage{graphicx}
\usepackage{multirow}
\usepackage{threeparttable}
\usepackage{bigstrut}
\usepackage{url}
\usepackage{txfonts}
\usepackage{color}

\usepackage{natbib}
\bibpunct{(}{)}{;}{a}{}{,}

\begin{document}
   \title{ynogkm: A New Public Code For Calculating time-like Geodesics In The Kerr-Newmann Spacetime}

   \author{Xiao-Lin. Yang
          \inst{1,2,3}  \and
           Jian-Cheng. Wang
          \inst{2,3}
          }

   \institute{Yunnan Astronomical Observatory, Chinese Academy of Sciences, Kunming 650011, P.R. China\
         \and
              Key Laboratory for the Structure and Evolution of Celestial Objects, Chinese Academy of Sciences, Kunming 650011, P.R. China\
         \and
              University of Chinese Academy of Sciences, Beijing 100049, P.R.
              China\
            }

\abstract{%
In this paper we present a new public code, named $ynogkm$, for the
fast calculation of time-like geodesics in the Kerr-Newmann (K-N)
spacetime, which is a direct extension of $ynogk$ calculating null
geodesics in a Kerr spacetime. Following the strategies used in
$ynogk$, we also solve the equations of motion analytically and
semi-analytically by using Weierstrass' and Jacobi's elliptic
functions and integrals, in which the Boyer-Lidquist (B-L)
coordinates $r$, $\theta$, $\phi$, $t$ and the proper time $\sigma$
are expressed as functions of an independent variable $p$ (Mino time). All of
the elliptic integrals are computed by Carlson's elliptic integral
method, which guarantees the fast speed of the code. Finally
the code is applied to a couple of toy problems.}


   \authorrunning{X. L. Yang and J. C. Wang}

   \titlerunning{ynogkm: a new public code}

   \maketitle
%

\section{Introduction}
\label{sec:introduction}

In the vicinity of the black hole and any other compact objects, the
gravitational field is extremely strong and the spacetime is
significant warped and twisted. Thus the general relativity effects
can not be ignored. The motion of free photons and test
particles in this curved spacetime is along geodesics if we do not
consider the external forces or perturbations exerting on them. The
assumption that photons and particles propagate along geodesic
trajectories is valid in most astrophysics contexts. The fast
calculation of the null and time-like geodesics in curved spacetime
is significantly important and has been
widely used in the Astrophysical researches (e.g.,
\cite{cunningham1973,Luminet1979,rauch1994,hackmann2010}).

The calculation and applications of null geodesics in a curved
spacetime, especially in a Kerr spacetime, have been discussed by
many authors in different attempts to date
(\cite{dexagol2009,hackmann2010,hackmann2013,
chan2013,yangwang2012}, and the references therein). To compute the
geodesics one can integrate a set of second-order differential
equations in any relativistic spacetime directly, or evaluate a set
of elliptic integrals of motion in a K-N spacetime. In the present
paper we focus on the latter approach. There
are four constants for any geodesic motions in a K-N spacetime
\citep{carter68}, which makes the
reduction of the order of motion equations possible.

To get the optical appearance of a star orbiting around an extreme Kerr
black hole, \cite{cunningham1973} calculated the null geodesics in
a Kerr spacetime based on the elliptic integral
method and proposed the impact parameters for the first time.
After that a method called ray-tracing was developed (e.g., \cite{Luminet1979}).
\cite{rauch1994} researched the optical caustics in a Kerr spacetime with
an attempt to explain rapid X-ray variability in AGN. As a gift they presented,
in tabular form, cases need to be considered for the calculation of both the null and time-like
geodesics in a Kerr spacetime. Similar discussions and results
are also given by \cite{li2005} in their Appendix.

The cases discussed by the above authors are very detailed but also very
complicated. As discussed in \cite{yangwang2012}, this sophisticated
situation can be significantly simplified by the introductions of
the Mino time $p$ \citep{mino2003} and the Weierstrass' elliptic integrals and functions
(also see \cite{hackmann2010,hackmann2013}, in which how to solve the
equations of geodesic motion in a more general instead of restricting
to the Kerr or K-N spacetime are discussed systematically by Mino time and all kinds of
elliptic functions). The Carlson's elliptic approach is quite suitable
and efficient for evaluating elliptic integrals and functions, which has
been demonstrated by \cite{dexagol2009} and \cite{yangwang2012}.

Motivated by the above discussions and the fact that there is no a
public code available in the present time to calculate time-like geodesics
in a K-N spacetimes for all coordinates (including the proper times)
at the same time, we extend the scheme of \cite{yangwang2012} from
null geodesics in a Kerr spacetime to time-like geodesics in a
K-N spacetime in this paper. As a result a new public code, named $ynogkm$ (Yun-Nan
Observatory Geodesic in a Kerr-Newmann spacetime for Massive
particles) is developed.

Analogous to $ynogk$, in $ynogkm$ we also express the B-L
coordinates $r$, $\theta$, $\phi$, $t$ and the proper time $\sigma$
as functions of the Mino time $p$ semianalytically by
using Weierstrass' and Jacobi's elliptic functions and integrals.
Such treatment makes the practical applications to be handled conveniently.
The Mino time $p$ is an integral value along a particular geodesic. All of the
elliptical integrals are computed by Carlson's approach. With a
similar way to $ynogk$, we also discuss how to compute the constants
of motion from the initial conditions, i.e., the initial
four-momentum of the particles measured under the local nonrotating
frame (LNRF, \cite{bardeen1972}). For a massive particle with
electric charge in a K-N spacetime, the number of constants
of motion becomes 4. For a photon whose rest mass $\mu_m$ and
electric charge $\epsilon$ are both zero, the number of constants of
motion is 2. When taking $\mu_m$ and $\epsilon$ to be zero, the
discussions here are reduced to those for null geodesics.

The paper is organized as follows. In section \ref{geoequ} we give
the equations of motion for an electric charged massive particle in
a K-N spacetime. In section
\ref{functions} we discuss the expressions of the B-L coordinates
and proper time as functions of parameter $p$. Then we reduce all of
the elliptic integrals to standard forms which are evaluated by
Carlson's approach. Next we discuss the calculation of constants of
motion from initial conditions in section \ref{motioncon}. A brief
introduction and discussion about the code are given in section
\ref{codeintroduction}. In section \ref{protest} we demonstrate the
applications of our code to toy problems in the literature.
Finally a brief summary is presented in section \ref{discconc}.
Throughout this paper the natural unit is used, in which the
constants G=c=1. The mass of central black hole M is also taken to
be 1, unless otherwise stated.

\section{The equations of motion for time-like geodesics}
\label{geoequ} We assume that the spin and electric charge of the
black hole are $a$ and $e$ respectively. Using the notation of
\citet{bardeen1972}, we can write the Kerr-Newman metric under the
B-L coordinate as
\begin{equation}
\begin{aligned}
ds^2=-e^{2\nu}dt^2+e^{2\psi}(d\phi-\omega
dt)^2+e^{2\mu_1}dr^2+e^{2\mu_2}d\theta^2,
\end{aligned}
\end{equation}
where
\begin{equation}
\begin{aligned}
&e^{2\nu}=\frac{\Sigma\Delta}{A}, \quad e^{2\psi}=\frac{\sin^2\theta
A}{\Sigma}, \quad e^{2\mu_1}=\frac{\Sigma}{\Delta}, \quad \\
&e^{2\mu_2}=\Sigma, \quad \omega=\frac{(2r-e^2)a}{A},
\end{aligned}
\end{equation}
and
\begin{equation}
\begin{aligned}
&\Delta=r^2-2r+a^2+e^2,\quad \Sigma=r^2+a^2\cos^2\theta, \quad \\
&A=(r^2+a^2)^2-\Delta a^2\sin^2\theta.
\end{aligned}
\end{equation}
\cite{carter68} gave the first-order differential equations
of motion for electric charged massive particles as follows:
\begin{eqnarray}
\label{defr}\Sigma \frac{dr}{d\lambda}&=&\pm\sqrt{R_r},\\
\label{deftheta}\Sigma\frac{d\theta}{d\lambda}&=&\pm\sqrt{\Theta_\theta},\\
\label{defphi}\Sigma\frac{d\phi}{d\lambda}&=&-(aE-\frac{L}{\sin^2\theta})+\frac{aT}{\Delta},\\
\label{deft}\Sigma\frac{dt}{d\lambda}&=&-a(aE\sin^2\theta-L)+\frac{(r^2+a^2)T}{\Delta},
\end{eqnarray}
where
\begin{eqnarray}\label{Rr}
&& T=E(r^2+a^2)-L a+e\epsilon r,\\
&& R_r=T^2-\Delta[\mu_m^2r^2+(L-aE)^2+Q],\\
&& \Theta_\theta=Q-\cos^2\theta[a^2(\mu_m^2-E^2)+L^2/\sin^2\theta],
\end{eqnarray}
and $\lambda=\tau/\mu_m$, $\tau$ is the proper time, $\mu_m$ is the
rest mass of the particle, $Q$ is the Carter constant, $E$ is the
energy tested by an observer at infinity, $L$ is the angular
momentum of the particle about the black hole spin axis, and
$\epsilon$ is the electric charge of the particle. From Equations
(\ref{defr})-(\ref{deft}) we can obtain the expression of the
four-momentum for a particle
\begin{eqnarray}
\label{fourmomentum} p_\mu = g_{\mu\nu}\frac{dx^\nu}{d\lambda}=
(-E,\pm\frac{\sqrt{R_r}}{\Delta},\pm\sqrt{\Theta_\theta},L).
\end{eqnarray}

Equivalently the equations of motion with integral forms can be
written as:
\begin{eqnarray}
\label{intrt}&&\pm\int^\theta \frac{d\theta}{\sqrt{\Theta_\theta}} = \pm\int^r\frac{dr}{\sqrt{R_r}},\\
\label{inttheta}&&\sigma = \frac{\tau E}{\mu_m} = \int^\theta
\frac{E a^2\cos^2\theta}{\sqrt{\Theta_\theta}}d\theta
+ \int^r\frac{E r^2}{\sqrt{R_r}}dr,\\
\label{intt}&&t = \sigma + 2\int^r\frac{N_r}{\Delta\sqrt{R_r}}dr,\\
\label{intphi}&&\phi = \int^\theta
\frac{L\csc^2\theta}{\sqrt{\Theta_\theta}}d\theta +
a\int^r\frac{r(2E-e\epsilon)-(Ee^2+La)}{\Delta\sqrt{R_r}}dr,
\end{eqnarray}
where
\begin{eqnarray}
\begin{aligned}
N_r=(2E+e\epsilon)r^3-Ee^2r^2+&[2a(Ea-L)\\
&+a^2e\epsilon]r-e^2a(Ea-L).
\end{aligned}
\end{eqnarray}
Here $\sigma$ is a new variable, which is related to the proper time
$\tau$ of the particle. For a photon, it becomes an affine parameter.

In many cases we only need the equations of motion with integral
forms. But in two special cases, i.e., the equatorial plane motion
and the spherical motion, we need the differential equations of
motion. In the former case, the particle is confined in the
equatorial plane, one has $Q=0$, $\theta\equiv \pi/2$, and thus
$\Theta_\theta\equiv 0$. Then the equations of motion with integral
forms become invalid, since $\Theta_\theta\equiv0$ appears in the
denominator. But from the differential equations we can get the right
equations to describe the plane motion. From Equation (\ref{defr}),
we have
\begin{eqnarray}
\begin{aligned}
&&\sigma = \lambda E = \int^r\frac{E r^2}{\sqrt{R_r}}dr.
\end{aligned}
\end{eqnarray}
Dividing both sides of Equation (\ref{defr}) by Equation
(\ref{deft}), we get
\begin{eqnarray}
&&t = \sigma + \int^r\frac{N_r}{\Delta\sqrt{R_r}}dr.
\end{eqnarray}
Similarly, from Equations (\ref{defr}) and (\ref{defphi}) we get
\begin{eqnarray}
&&\phi = L\int^r \frac{dr}{\sqrt{R_r}} +
a\int^r\frac{r(2E-e\epsilon)-(Ee^2+La)}{\Delta\sqrt{R_r}}dr.
\end{eqnarray}

For spherical motion, we have $R_r\equiv0$, thus the equations of
motion with integral forms become invalid, because $R_r$ appears in
the denominator. Similarly, from Equation (\ref{deftheta}) we obtain
\begin{eqnarray}
\sigma = E\int d\lambda =  E a^2\int^\theta
\frac{\cos^2\theta}{\sqrt{\Theta_\theta}}d\theta + E
r^2\int^\theta\frac{d\theta}{\sqrt{\Theta_\theta}}.
\end{eqnarray}
From Equations (\ref{deftheta}) and (\ref{deft}) we get
\begin{eqnarray}
t = \sigma + \frac{N_r}{\Delta}\int^\theta\frac{d\theta}{
\sqrt{\Theta_\theta}}.
\end{eqnarray}
And from Equations (\ref{deftheta}) and (\ref{defphi}) we get
\begin{eqnarray}
\label{intphic}\phi = L\int^\theta
\frac{\csc^2\theta}{\sqrt{\Theta_\theta}}d\theta +
a\frac{r(2E-e\epsilon)-(Ee^2+La)}{\Delta}\int^\theta\frac{d\theta}{\sqrt{\Theta_\theta}}.
\end{eqnarray}
With Equations (\ref{intrt})-(\ref{intphic}), we can calculate the
geodesics by evaluating the elliptical integrals, instead of solving
the differential equations of motion, and can also express the B-L
coordinates and proper time as functions of a parameter $p$. In the
next section we discuss how to get these functions semianalytically
by elliptical functions and integrals.

\section{The expressions of B-L coordinates and proper time as functions of $p$}
\label{functions}
\subsection{The turning points}
As discussed in \cite{yangwang2012}, when we introduce a new
parameter $p$ with following definition from Equation (\ref{intrt})
\begin{eqnarray}
\label{defp1} p = \pm \int^r\frac{ dr}{\sqrt{R}} =  \pm
\int^\mu\frac{ d\mu}{\sqrt{\Theta_{\mu}}},
\end{eqnarray}
we can get functions $r(p), \mu(p), \phi(p), t(p),$ and $\sigma(p)$
by the equations of motion with integral forms, where $\mu=\cos\theta$ and
\begin{eqnarray}
\begin{aligned}
 R(r) = &\frac{R_r}{E^2} = (1-m^2)r^4+2(m^2+e\varepsilon)r^3- [q+\lambda^2\\
 &+a^2(m^2-1)+e^2(m^2-\varepsilon^2)]r^2+ 2[q+(a-\lambda)^2\\
 &+ea\varepsilon(a-\lambda)]r-e^2(a-\lambda)^2-(a^2+e^2)q,
\end{aligned}\\
\begin{aligned}
\Theta_\mu(\mu) &= \frac{\Theta_\theta \sin^2\theta}{E^2}\\
&=a^2(m^2-1)\mu^4 - [q+\lambda^2+a^2(m^2-1)]\mu^2 + q.
\end{aligned}
\end{eqnarray}
Since the signs before the integrals are the same with $dr$ and
$d\theta$, the parameter $p$ monotonously increases along a particular
geodesic. Here $\lambda=L/E$, $q=Q/E^2$, $m=\mu_m /E$, and
$\varepsilon=\epsilon/E$, which are defined as constants of motion
throughout this paper. Because $R$ and $\Theta_\mu$ are quartic
(when $m\neq 1$) or cubic (when $m=1$) polynomials, the integrals
about $r$ and $\mu$ are elliptical integrals, which are reduced to the
Weierstrass' standard elliptic integrals (or
Legendre's ones when equation $R(r)=0$ has no real roots).

Since both $R$ and $\Theta_\theta$ appear under the radical sign in
the equations of motion, they must be nonnegative. The critical points
satisfying $R(r)=0$ or $\Theta_\theta(\theta)=0$ are so-called
turning points, in which the corresponding coordinate velocity is
zero. When the motion of a particle is bounded in $r$ or in
$\theta$ coordinate, two turning points exist for the coordinate. We
use $r_{\mathrm{tp}_1}, r_{\mathrm{tp}_2}$ and
$\theta_{\mathrm{tp}_1}$, $\theta_{\mathrm{tp}_2}$ to denote the
coordinates of these points, and assume that $r_{\mathrm{tp}_1}\leq
r_{\mathrm{tp}_2}$,
$\theta_{\mathrm{tp}_1}\leq\theta_{\mathrm{tp}_2}$. Since
$p_r=\pm\sqrt{R_r}/\Delta, p_\theta =\pm \sqrt{\Theta_\theta}$, when
$p_r=0$ or $p_\theta=0$ at the initial point, we have
$R_r(r_{\mathrm{ini}}) = 0, \Theta_\theta(\theta_{\mathrm{ini}})=0$,
implying that the initial point is a turning point and
$r_{\mathrm{ini}}$ (or $\theta_{\mathrm{ini}}$) is equal to one of
$r_{\mathrm{tp}_1}, r_{\mathrm{tp}_2}$ (or $\theta_{\mathrm{tp}_1}$,
$\theta_{\mathrm{tp}_2}$). Then we have $r_{\mathrm{ini}}\in
[r_{\mathrm{tp}_1}, r_{\mathrm{tp}_2}]$ and $\theta_{\mathrm{ini}}
\in [\theta_{\mathrm{tp}_1}, \theta_{\mathrm{tp}_2}]$.

When $r_{\mathrm{tp}_2}$ does not exist at all (or equivalently,
$r_{\mathrm{tp}_2}=\infty$) and $r_{\mathrm{tp}_1}> r_\mathrm{h}$,
the particle will eventually goto infinity far away. When
$r_{\mathrm{tp}_1}$ does not exist (or $r_{\mathrm{tp}_1}<
r_{\mathrm{h}}$) and $r_{\mathrm{tp}_2}>r_{\mathrm{h}}$, then the
particle will eventually fall into the event horizon of the black
hole. If (1) both $r_{\mathrm{tp}_1}$ and $r_{\mathrm{tp}_2}$ do not
exist, this case equivalently corresponds to that the equation $R(r)=0$ has no
real roots; or (2) $r_{\mathrm{tp}_2}$ does not exist and
$r_{\mathrm{tp}_1}$ exists but $r_{\mathrm{tp}_1}< r_{\mathrm{h}}$,
for the both cases the particle can move from infinity to the event horizon freely.

To get the $\theta$ coordinate of a turning point, we usually solve
the equation $\Theta_\mu(\mu) = 0$ to get $\mu_\mathrm{tp}$
(=$\cos\theta_{\mathrm{tp}}$) instead of solving the equation
$\Theta_\theta(\theta)=0$. The roots of two equations are exactly the same
except these special cases with constant $\lambda=0$. The equation
$\Theta_\mu(\mu)=0$ with $\lambda=0$ has real roots $\pm1$, or $0,
\pi$, which are not the roots of equation $\Theta_\theta(\theta)=0$,
indicating that a particle with $\lambda=0$ can move from $0$ to
$\pi$ freely and can go through the spin axis due to non-zero
poloidal velocity $p_\theta=\pm\sqrt{\Theta_\theta}$ at the spin
axis. Meanwhile, the particle changes the sign of its angular
velocity $d\theta/d\lambda$, and its azimuthal coordinate jumps from
$\phi$ to $\phi\pm\pi$ \citep{shakura1987} instantaneously.

\subsection{$\mu$ and $r$ coordinates}
In this section we express $\mu$ and $r$ as functions of
parameter $p$, i.e., $\mu=\mu(p), r=r(p)$. The procedure to get
these explicit expressions for electric charged
massive particles is quite tedious but similar to the procedure
for photons (one can refer to the discussions in \cite{yangwang2012}).
Thus there is no need to present the details of the procedure. For
the purpose of easier referring we present expressions of $\mu(p), r(p)$
in tabular form. See table \ref{table1}.

For $r$, there are five cases:\\
1. $m\neq 1$ and equation $R(r)=0$ has one real root at least.\\
2. $1-m^2>0$ (or $|E|>\mu_m$) and $R(r)=0$ has no real roots.\\
3. $1-m^2<0$ (or $|E|<\mu_m$) and $R(r)=0$ has no real roots.\\
4. $|m|=1$ (or $|E|=\mu_m$) and the geodesic is unbounded.\\
5. $|m|=1$ (or $|E|=\mu_m$) and the geodesic is bounded.

\newcommand{\ZZ}[2]{\rule[#1]{0pt}{#2}}
\newcommand{\MC}[3]{\multicolumn{#1}{#2}{#3}}
\newcommand{\rms}[1]{\mathrm{#1}}
\newcommand{\lp}{\left(}
\newcommand{\rp}{\right)}
\newcommand{\lf}{\left[}
\newcommand{\rf}{\right]}

\begin{table}
\label{table1}
\begin{center}
    \begin{threeparttable}
\begin{tabular}{c|l}
  \MC{2}{c}{\textbf{Table} 1. \; Expression of $\mu(p)$} \\
\hline\hline
    \ZZ{-5pt}{15pt} Case &  \MC{1}{c}{$\mu(p)$}     \\
  \hline
    \multirow{4}{*}{} $a\neq 0$  & \begin{minipage}[b][\height]{20em}
                                    \begin{eqnarray}
                                        \displaystyle \mu(p)=\frac{b_0}
                                        {4\wp(p+\Pi_\mu;g_2,g_3)-b_1}+\mu_{\mathrm{tp}_1},
                                    \end{eqnarray}
                                 \end{minipage} \\
                                 \cline{2-2}
       $m\neq 1$ & \ZZ{-5pt}{15pt}$b_0 = 4a^2(m^2-1)\mu_{\mathrm{tp}_1}^3
                               -2[q+\lambda^2+a^2(m^2-1)]\mu_{\mathrm{tp}_1},$ \\
                 & \ZZ{-5pt}{15pt}$b_1 =  2a^2(m^2-1)\mu_{\mathrm{tp}_1}^2-\frac{1}{3}[q+\lambda^2+a^2(m^2-1)], $\\
                 & \MC{1}{l}{$b_2$ =  $\frac{4}{3}a^2(m^2-1)\mu_{\mathrm{tp}_1}, \quad b_3$ =  $a^2(m^2-1), $}\; $g_2,\;g_3\tnote{1}.$ \\
    \hline
       \hline
    \ZZ{-5pt}{15pt} Case &  \MC{1}{c}{$\Pi_\mu$}     \\
  \hline
    \multirow{1}{*}{} $a\neq 0$  & \ZZ{-5pt}{15pt} $|\Pi_\mu|=|\wp^{-1}[z(\mu_{\rms{ini}});g_2,g_3]|$,
                    \;\;$z(\mu)=\frac{b_0}{4}\frac{1}{(\mu-\mu_{\mathrm{tp}_{1}})}+\frac{b_1}{4},$  \\
       $m\neq 1$ &  \begin{minipage}[b][\height]{20em}
                    \begin{eqnarray*}
                       \Pi_\mu\left\{
                           \begin{array}{ll}
                               >0, \quad & p_\theta>0, \\
                               \left.
                               \begin{array}{l}
                               =\pm n\omega'\tnote{2}, \quad\;\quad\quad \theta_{\rms{ini}} = \theta_{\rms{tp_1}} \\
                               =\pm (\frac{1}{2}+n)\omega', \quad \theta_{\rms{ini}} = \theta_{\rms{tp_2}}
                               \end{array}
                               \right\} &p_\theta=0, \\
                               <0, \quad &p_\theta<0.
                           \end{array}
                       \right.
                    \end{eqnarray*}
                    \end{minipage} \\
                 \hline
     \hline
\end{tabular}
\begin{tablenotes}
  \footnotesize
  \item[1]  $g_2 = \frac{3}{4}(b_1^2-b_0b_2), $  $g_3 = \frac{1}{16}(3b_0b_1b_2-2b_1^3-b_0^2b_3)$.
  \item[2]  where $\omega'$ is the real period of $\wp(z;g_2,g_3)$ and $n=0, 1, 2, ...$
\end{tablenotes}
\end{threeparttable}
\end{center}
\end{table}

\begin{table}
\label{table2}
\begin{center}
    \begin{threeparttable}
\begin{tabular}{c|l}
  \MC{2}{c}{\textbf{Table} 2.\; Expression of $r(p)$} \\
  \hline \hline
    \ZZ{-5pt}{15pt} Case &  \MC{1}{c}{$r(p)$}     \\
  \hline
   1  &  \begin{minipage}[b]{22em} 
             \begin{eqnarray}
                 \displaystyle r(p)= \frac{b_0}{4\wp(p+\Pi_r;g_2,g_3)-b_1}+r_{\mathrm{tp}_{1}},
             \end{eqnarray}
         \end{minipage}
            \\ \hline
  \begin{minipage}[b]{1em}
             \begin{eqnarray*}
               \\
               2\\
               \\
             \end{eqnarray*}
   \end{minipage}   &   \begin{minipage}[b][\height]{22em} 
            \begin{eqnarray*} 
                \begin{array}{l} 
                        \displaystyle r_{\pm}(p) = v\displaystyle +\frac{-(u-v)\pm s(\lambda_1-\lambda_2)z(p)
                        \sqrt{1-z(p)^2}}{(\lambda_1-\lambda_2)z(p)^2-(\lambda_1-1)},  \\
                        z(p)=\mathrm{sn}(s\sqrt{\lambda_1(1-m^2)}p\mp \Pi_r|k^2),
                \end{array}  
            \end{eqnarray*} 
            \begin{eqnarray} 
                \label{sign_r1}
                \begin{array}{l} 
                    r(p)=\left\{\begin{array}{ll}
                       r_-(p), & p_r>0,\\
                       r_+(p), & p_r<0.
                    \end{array}\right.
                \end{array}  
           \end{eqnarray} 
         \end{minipage}
            \\
          \hline
   \begin{minipage}[b]{1em}
             \begin{eqnarray*}
               \\
               3\\
               \\
             \end{eqnarray*}
   \end{minipage}    &   \begin{minipage}[b][\height]{22em} 
            \begin{eqnarray*} 
                \begin{array}{l} 
                       \displaystyle r_{\pm}(p) = u\\
                       \displaystyle  +\frac{-\lambda_1(u-v)/(\lambda_1+1)\pm wz(p)
                       \sqrt{1-z(p)^2[(1+\lambda_1)/k^2]^2}} {1-z(p)^2[(1+\lambda_1)/k^2]},\\
                       z(p)=\mathrm{sn}(s\sqrt{\lambda_2(1-m^2)}p\mp \Pi_r|k^2),
                \end{array}
           \end{eqnarray*} 
            \begin{eqnarray} 
                \label{sign_r2}
                \begin{array}{l} 
                       r(p)=\left\{\begin{array}{ll}
                       r_-(p), & p_r>0,\\
                       r_+(p), & p_r<0.
                    \end{array}\right.
                \end{array}  
           \end{eqnarray} 
         \end{minipage}
           \\
          \hline
    4 & \begin{minipage}[b][\height]{22em}
                           \begin{eqnarray}
                                 \label{rp4}
                                 \displaystyle r(p) = \frac{1}{b_0}[4\wp(p+\Pi_r;g_2,g_3)-b_1],
                           \end{eqnarray}
                         \end{minipage}   \\
                       \hline
    5                   &  \begin{minipage}[b][\height]{22em}
                           \begin{eqnarray}
                                 \displaystyle  r(p) = \frac{1}{b_0}\left[4e_2-b_1-\frac{4(e_1-e_2)(e_2-e_3)}
                                                 {\wp(p+\Pi_\xi;g_2,g_3)-e_2}\right].
                           \end{eqnarray}
                         \end{minipage}   \\
  \hline \hline
\end{tabular}
\end{threeparttable}
\end{center}
\end{table}

\begin{table*}
\label{table3}
\begin{center}
    \begin{threeparttable}
\begin{tabular}{c|l|l}
  \MC{3}{c}{\textbf{Table} 3.\; Definitions of Table 2.} \\
  \hline \hline
    \ZZ{-5pt}{15pt} Case &  \MC{1}{c}{$b_0,b_1,b_2,b_3,g_2,g_3,\lambda_1,\lambda_2,k^2$} \vline & \MC{1}{c}{$\Pi_r, \Pi_\xi$}    \\
  \hline
   \begin{minipage}[b]{1em}
             \begin{eqnarray*}
             1
             \end{eqnarray*}
   \end{minipage}  &  \begin{minipage}[b]{27em} 
             \begin{eqnarray*}
             \begin{array}{lll}
                 b_0 &=& 4(1-m^2)r_{\mathrm{tp}_{1}}^3+6(m^2+e\varepsilon) r_{\mathrm{tp}_{1}}^2-2[q+\lambda^2+a^2(m^2-1)\nonumber \\
                 &&+e^2(m^2-\varepsilon^2)]r_{\mathrm{tp}_{1}}+2[q+(\lambda-a)^2+ae\varepsilon(a-\lambda)], \\
                 b_1 &=& 2(1-m^2)r_{\mathrm{tp}_{1}}^2+2(m^2+e\varepsilon)
                 r_{\mathrm{tp}_{1}}-\frac{1}{3}[q+\lambda^2 \nonumber\\
                 &&+a^2(m^2-1)+e^2(m^2-\varepsilon^2)]r_{\mathrm{tp}_{1}},\\
                 b_2 &=& \frac{4}{3}(1-m^2)r_{\mathrm{tp}_{1}} + \frac{2}{3}(m^2+e\varepsilon),\quad  b_3 = 1-m^2, \;\;\;\;g_2,\;\;g_3\tnote{1}\\
             \end{array}
             \end{eqnarray*}
         \end{minipage}  &  \begin{minipage}[b]{19em}
                              \begin{eqnarray*}
                              \begin{array}{lll}
                                 |\Pi_r|=|\wp^{-1}[z(r_{\mathrm{ini}});g_2,g_3]|,\;\;
                                        z(r)=\frac{b_0}{4}\frac{1}{(r-r_{\mathrm{tp}_{1}})}+\frac{b_1}{4}, \\
                                 \Pi_r= \left\{
                                 \begin{array}{ll}
                                 |\Pi_r|,& p_r>0,\\
                                \left. \begin{array}{lr}
                                      \pm n\omega'\tnote{1}, &\; r_{\rms{ini}} = r_{\rms{tp}_1}\\
                                      \pm(\frac{1}{2}+n)\omega',&\; r_{\rms{ini}} = r_{\rms{tp}_2}\\
                                      \end{array}
                                  \right\} &p_r=0, \\
                                 -|\Pi_r|, &p_r<0,
                                 \end{array}
                                 \right. \\
                            \end{array}
                            \end{eqnarray*}
                            \end{minipage}      \\
          \hline
      \begin{minipage}[b]{1em}
             \begin{eqnarray*}
             \begin{array}{c}
               2
             \end{array}
             \end{eqnarray*}
   \end{minipage}  &   \begin{minipage}[b][\height]{27em} 
            \begin{eqnarray} 
                \lambda_{1,2}=\frac{(u-v)^2+w^2+s^2\pm\sqrt{[w^2+s^2+(u-v)^2]^2-4s^2w^2}}{2s^2},
           \end{eqnarray} 
           \begin{eqnarray*}
                \lambda_1>1>\lambda_2>0, \quad k^2=\frac{\lambda_1-\lambda_2}{\lambda_1},
           \end{eqnarray*}
         \end{minipage} &   \begin{minipage}[b]{18em}
                              \begin{eqnarray*}
                                     \Pi_r=|\mathrm{sn}^{-1}[z(r_{\mathrm{ini}})|k^2]|,\quad \alpha_1 = \frac{\lambda_1v-u}{\lambda_1-1},
                             \end{eqnarray*}
                             \begin{eqnarray}
                                 \label{tr1}z(r)=\sqrt{\frac{\lambda_1-1}{\lambda_1-\lambda_2}}\frac{(r-\alpha_1)}{\sqrt{(r-v)^2+s^2}},
                             \end{eqnarray}
                            \end{minipage}  
            \\
          \hline
   \begin{minipage}[b]{1em}
             \begin{eqnarray*}
                3
             \end{eqnarray*}
   \end{minipage}   &   \begin{minipage}[b][\height]{27em} 
            \begin{eqnarray} 
               \lambda_{1,2}=\frac{-(u-v)^2-w^2-s^2\pm\sqrt{[w^2+s^2+(u-v)^2]^2-4s^2w^2}}{2s^2},
           \end{eqnarray} 
           \begin{eqnarray*}
                0>\lambda_1>-1>\lambda_2, \quad k^2=\frac{\lambda_2-\lambda_1}{\lambda_2},
           \end{eqnarray*}
         \end{minipage}  &                  \begin{minipage}[b]{18em}
                                            \begin{eqnarray*}
                                               \Pi_r=|\mathrm{sn}^{-1}[z(r_{\mathrm{ini}})|k^2]|,\quad \alpha_1=\frac{\lambda_1v+u}{\lambda_1+1},
                                            \end{eqnarray*}
                                            \begin{eqnarray}
                                                 \label{tr2}z(r)=\sqrt{\frac{\lambda_1-\lambda_2}{-\lambda_2
                                                 (\lambda_1+1)}}\frac{(r-\alpha_1)}{\sqrt{(r-u)^2+w^2}},
                                            \end{eqnarray}
                                            \end{minipage}
           \\
          \hline
   4 & \begin{minipage}[b]{27em}
                           \begin{eqnarray*}
                           \begin{array}{lrl}
                               b_0 = 2(1+e\varepsilon),& &b_1 = -\frac{1}{3}[q+\lambda^2+e^2(1-\varepsilon^2)],\\
                               b_2 = \frac{2}{3}[q+(a-\lambda)^2+ea\varepsilon(a-\lambda)],& &b_3 = -e^2(a-\lambda)^2-(a^2+e^2)q,\\
                               g_2,\quad g_3\tnote{1} 
                           \end{array}
                           \end{eqnarray*}
                         \end{minipage} &  \begin{minipage}[b]{18em}
                                             \begin{eqnarray*}
                                                \begin{array}{l}
                                                  |\Pi_r| =|\wp^{-1}[z(r_{\mathrm{ini}});g_2,g_3]| ,\quad z(r) = \frac{b_0}{4}r+\frac{b_1}{4},    \\
                                             \Pi_r=\left\{
                                             \begin{array}{lr}
                                             -\mathrm{sign}(b_0)|\Pi_r|,&\quad p_r>0,\\
                                             \pm(\frac{1}{2}+n)\omega'\tnote{1}, \;\;r_{\rms{ini}} = r_{\rms{tp}_1}, & p_r=0,\\
                                             \mathrm{sign}(b_0)|\Pi_r|,&\quad p_r<0,
                                             \end{array}
                                             \right.
                                              \end{array}
                                             \end{eqnarray*}
                                           \end{minipage}  \\
                       \hline
    5                   &  \begin{minipage}[b]{27em}
                               \begin{eqnarray*}
                               \begin{array}{lrl}
                                   b_0 = 2(1+e\varepsilon),& &b_1 = -\frac{1}{3}[q+\lambda^2+e^2(1-\varepsilon^2)],\\
                                   b_2 = \frac{2}{3}[q+(a-\lambda)^2+ea\varepsilon(a-\lambda)],& &b_3 = -e^2(a-\lambda)^2-(a^2+e^2)q,\\
                                   g_2,\quad g_3\tnote{1} 
                               \end{array}
                               \end{eqnarray*}
                                $e_1,e_2,e_3$  are three real roots of  $4z^3-g_2z-g_3=0$ and $e_3<e_2<e_1$.
                         \end{minipage}  &   \begin{minipage}[b]{20em}
                                             \begin{eqnarray*}
                                             \begin{array}{lll}
                                                   |\Pi_\xi|=|\wp^{-1}[\xi(r_{\mathrm{ini}});g_2,g_3]|,\;\;\;\\
                                                   \displaystyle \xi(r) = e_2-\frac{(e_1-e_2)(e_2-e_3)}{z(r)-e_2},
                                                   \quad z(r) = \frac{b_0}{4}r+\frac{b_1}{4},  \\
                                             \Pi_\xi=\left\{
                                             \begin{array}{lr}
                                             -\mathrm{sign}(b_0)|\Pi_\xi|,&\; p_r>0,\\
                                                \left.\begin{array}{lr}
                                                     \pm(\frac{1}{2}+n)\omega'\tnote{1}, \;\;&z(r_{\rms{ini}}) = e_3, \\
                                                     \pm n\omega', \;\;&         z(r_{\rms{ini}}) = e_2, \\
                                                \end{array} \right\}\; & p_r=0,\\
                                             \mathrm{sign}(b_0)|\Pi_\xi|,&\; p_r<0.
                                             \end{array}
                                             \right.
                                             \end{array}
                                             \end{eqnarray*}
                                           \end{minipage}  \\
                         \hline
  \hline
\end{tabular}
\begin{tablenotes}
  \footnotesize
  \item[1]  See the footnote of Table 1.
\end{tablenotes}
\end{threeparttable}
\end{center}
\end{table*}

\begin{table}
\label{table4}
\begin{center}
    \begin{threeparttable}
\begin{tabular}{c|l}
  \MC{2}{c}{\textbf{Table} 4. \; Expression of $p$} \\
  \hline \hline
    \ZZ{-5pt}{15pt} Case &  \MC{1}{c}{$p$}     \\
  \hline
   1,\;\; 4,\;\;5  &  \begin{minipage}[b]{20em} 
             \begin{eqnarray}
                 \displaystyle p=\int^z\frac{dz}{\sqrt{4z^3-g_2z-g_3}},
             \end{eqnarray}
         \end{minipage}
            \\ \hline
  \begin{minipage}[b]{1em}
             \begin{eqnarray*}
                  2
             \end{eqnarray*}
   \end{minipage}   &   \begin{minipage}[b][\height]{20em} 
            \begin{eqnarray} 
                        p=\int^{z}\frac{dz}{s\sqrt{\lambda_1(1-m^2)}\sqrt{(1-z^2)(1-k^2z^2)}},
            \end{eqnarray} 
         \end{minipage}
            \\
          \hline
   \begin{minipage}[b]{1em}
             \begin{eqnarray*}
               3
             \end{eqnarray*}
   \end{minipage}    &   \begin{minipage}[b][\height]{20em} 
            \begin{eqnarray} 
                 p=\int^{z}\frac{dz}{s\sqrt{\lambda_2(1-m^2)}\sqrt{(1-z^2)(1-k^2z^2)}}.
           \end{eqnarray} 
         \end{minipage}
           \\
  \hline \hline
\end{tabular}
\end{threeparttable}
\end{center}
\end{table}

\label{rcoordinate}
In case 1, $R(r)=0$ with $m\neq 1$ at least
has one real root. We do not care how many real roots the equation
has and their practical distribution. When the equation $R(r)=0$ has
real roots, $r_{\mathrm{tp}_1}$ (or $r_{\mathrm{tp}_2}$) does exist,
and can be easily determined with given
$r_{\mathrm{ini}}$\footnote{Actually $r_{\mathrm{tp}_1}$ exists for
all cases that we discussed in this paper.}.

In cases 2 and 3, $R(r)=0$ has no real roots,
but two pairs of complex conjugate roots written as:
\begin{eqnarray}
 r_1=u+iv, \quad r_2=u-iv,\\
 r_3=w+is, \quad r_4=w-is.
\end{eqnarray}
To avoid dealing with complex integral we use Jacobi's elliptic
function to express $r$ instead of Weierstrass' ones.

For cases 4 and 5, $1-m^2=0$, thus $R(r)$ reduces to
\begin{eqnarray}
\begin{aligned}
 R(r) =& 2(1+e\varepsilon)r^3-[q+\lambda^2+e^2(1-\varepsilon^2)]r^2+2[q+(a-\lambda)^2\\
 &+ea\varepsilon(a-\lambda)]r-e^2(a-\lambda)^2-(a^2+e^2)q,
\end{aligned}
\end{eqnarray}
 From the expression of $r(p)$ in Equation (\ref{rp4}), we
know that when $p\pm\Pi_r=\omega'$, $\wp(p\pm\Pi_r;g_2,g_3)=\infty$,
i.e., $r(p)=\infty$, meaning that no matter what initial value $p_r$
takes, the particle shall go to infinity eventually.
As a result, the particular geodesic is unbounded.

\subsection{$\phi$, $t$ coordinates and the proper time $\sigma$}
As discussed in \cite{yangwang2012}, the expressions of $\phi, t$,
and $\sigma$ as functions of parameter $p$ can be converted to
evaluate the elliptic integrals appeared in the equations of motion
with a given $p$. We divide the process into two steps. In the first
step, the path and limits of the integrals are determined. In the second step,
the integrals are reduced to standard forms, which are evaluated by Carlson's approach.

\subsubsection{The path and limits of integrals}
For convenience, we use $F_r$ and $F_\mu$ to represent the complicated
integrands in integrals of $r$ and $\mu$ respectively.

\begin{figure}
\centering
\includegraphics[width=0.4\textwidth,angle=0]{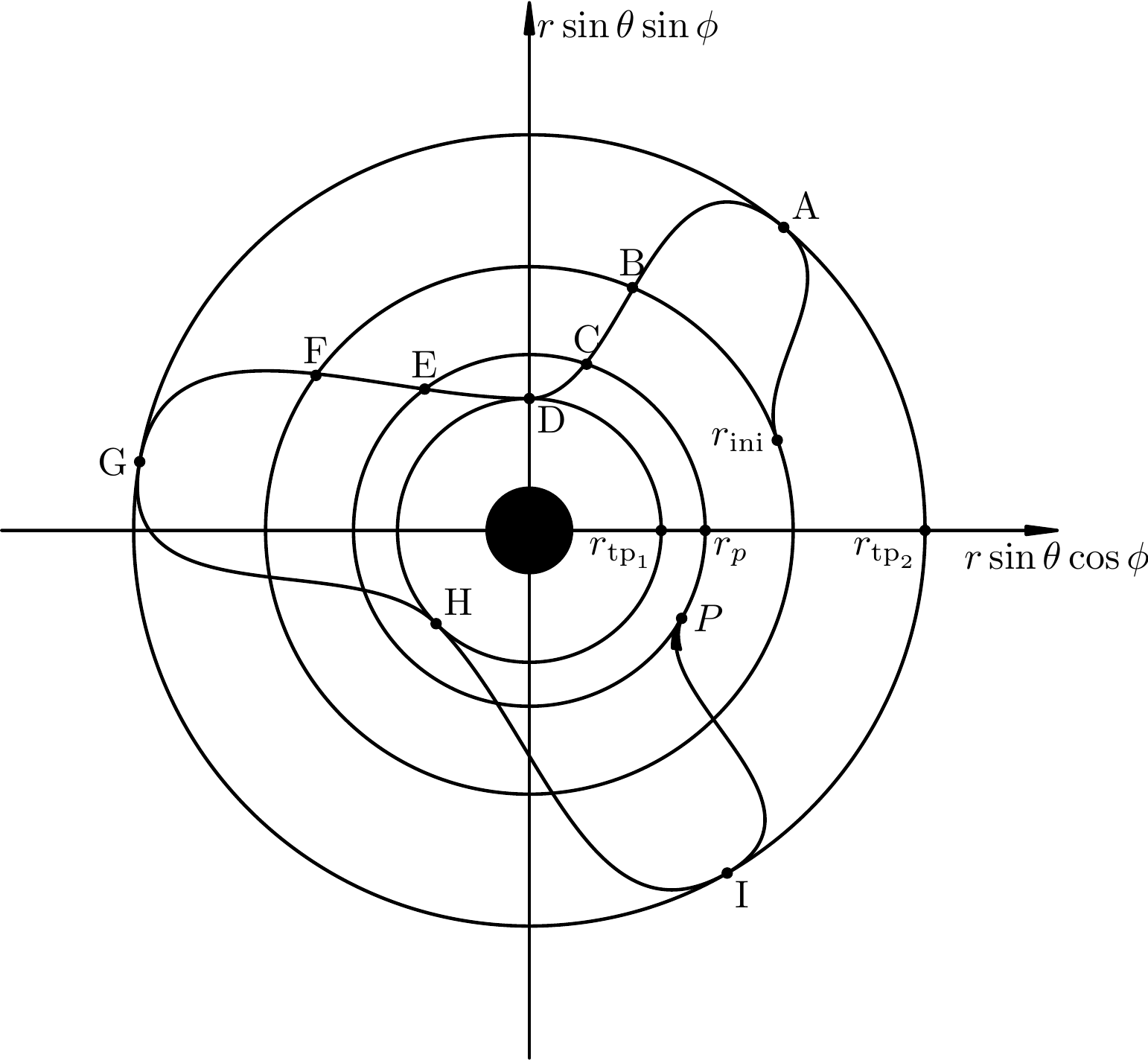}
\caption{  The motion of $r$ illustrated schematically. It has been
projected onto the equatorial plane of the black hole.
$r_{\mathrm{tp}_1}$ and $r_{\mathrm{tp}_2}$ are radial turning
points, in which the motion is confined. Point P indicates the
position for a given $p$. The whole integral path is not monotonous,
and should be divided into several sections, each one has the
maximum proper length. Such as DA, DG, HG, HI, PI. From the
definitions of $I_{0\alpha r}, I_{1\alpha r},$ and $I_{2\alpha r}$
(see text), one has $I_{0\alpha r}=\int^C_B=\int^E_F$, $I_{1\alpha
r}=\int^D_C=\int^D_E$, $I_{2\alpha r}=\int^C_A=\int^E_G$,
$IT_{1\alpha r}=\int^D_B=\int^D_F$, and $IT_{2\alpha
r}=\int^B_A=\int^F_G$, etc.. } \label{fig:rmotion}
\end{figure}

The path is not monotonic when one or more than one turning points
exist for $r$ and $\mu$. The path is divided into several parts, in
which each one has the maximum monotonic length, and the integrals
are the sum of all individual part. In Figure
\ref{fig:rmotion}, the integral path of $r$ coordinate for a
particular bounded geodesic is
illustrated schematically. The motion of the particle is confined
between two turning points, $r_{\mathrm{tp}_1}$ and
$r_{\mathrm{tp}_2}$.

There are four important points in a particular path for $r$ (or for $\mu$), they are related to the
integral limits. They are: 1. the initial position $r_{\rms{ini}}$
(or $\mu_{\rms{ini}}$); 2. the two turning points, $r_{\rms{tp}_1}$ and
$r_{\rms{tp}_2}$ (or $\mu_{\rms{tp}_1}$, $\mu_{\rms{tp}_2}$);
3. the position corresponding to a given $p$, $r_p$ and
$\mu_p$. The $z$ values of these points are
$z_{\rms{ini}}$, $z_{\rms{tp}_1}$ and $z_{\rms{tp}_2}$, and $z_p$.
The former three ones can be evaluated from $z(r)$ functions given in
the right column of Table 3. $z_p$ can be evaluated from function
$z=\wp(p\pm\Pi_r;g_2,g_3)$ for cases 1, 4, 5 and
$z=\mathrm{sn}(s\sqrt{\lambda_1(1-m^2)}p\pm\Pi_r|k^2)$ for case 2,
and $z=\mathrm{sn}(s\sqrt{\lambda_2(1-m^2)}p\pm\Pi_r|k^2)$ for
case 3 (for $\mu$, $z_p$ can be computed from
$z=\wp(p\pm\Pi_\mu;g_2,g_3)$). It is noted that the functions $z(r)$
are monotonously decreasing, we have $z_{\mathrm{tp}_1}\geq
z_{\mathrm{ini}}\geq z_{\mathrm{tp}_2}$ and $z_{\mathrm{tp}_1}\geq
z_{p}\geq z_{\mathrm{tp}_2}$ (for $\mu$, since the function $z(\mu)$
given in Table 1. is monotonously increasing, the two
relationships are still valid).

In addition to $z_p$, for a given $p$, we can also obtain the number of times that the
particle meets the two turning points. We assume that the particle
meets $r_{\rms{tp}_1}$ (or
$\mu_{\rms{tp}_1}$) for $Nt_1$ times, and $r_{\rms{tp}_2}$ (or
$\mu_{\rms{tp}_2}$) for $Nt_2$ times. $Nt_1$ or $Nt_2$ is zero if
$r_{\rms{tp}_1}$ or $r_{\rms{tp}_2}$ does not exist.
To get $Nt_1$ and $Nt_2$ for a given $p$, we define the
following five integrals with the help of Table 4:
\begin{eqnarray}
\begin{aligned}
p_0=\int^{z_p}_{z_{\mathrm{ini}}}&W(z) dz,\quad
p_1=\int^{z_{\mathrm{tp}_1}}_{z_p}W(z) dz,\quad
p_2=\int^{z_p}_{z_{\mathrm{tp}_2}}W(z) dz,\\
&I_1=\int^{z_{\mathrm{tp}_{1}}}_{z_{_{\mathrm{ini}}}}W(z) dz,\quad
I_2=\int_{z_{\mathrm{tp}_{2}}}^{z_{_{\mathrm{ini}}}}W(z) dz,
\end{aligned}
\end{eqnarray}
where $W(z)$ represents the integrands in Table 4. Apparently we have $p_1\ge0$
and $p_2\ge0$, and
\begin{eqnarray}
p_1=I_1-p_0,\quad p_2=I_2+p_0.
\end{eqnarray}
With the above definitions, we get the following identity
\begin{eqnarray}
\begin{aligned}
 \label{t1t2}p =& -\mathrm{sign}(p_\beta)p_0+2Nt_1p_1+2Nt_2p_2,\\
=&-[\mathrm{sign}(p_\beta)+2Nt_1-2Nt_2]p_0+2Nt_1I_1+2Nt_2I_2,
\end{aligned}
\end{eqnarray}
where $\beta=r$ or $\theta$, and $p_\beta$ is the initial value of
$r$ or $\theta$ component of the four-momentum. One can get $Nt_1$
and $Nt_2$ from the above equations by trial and error, because
$Nt_1$ and $Nt_2$ increase regularly as the particle moves, i.e.,
when $p_{\beta}>0$ (or, $p_\beta=0$ and
$z_{\rms{ini}}=z_{\rms{tp}_1}$), $Nt_1$ and $Nt_2$
increase as:
\begin{eqnarray}
     \nonumber Nt_1 = 0,\,\,\,0,\,\,\,1,\,\,\,1,\,\,\,2,\,\,\,2,\,\,\,3,\,\,\,3...\\
     \nonumber  Nt_2 = 0,\,\,\,1,\,\,\,1,\,\,\,2,\,\,\,2,\,\,\,3,\,\,\,3,\,\,\,4...
\end{eqnarray}
When $p_\beta<0$ (or, $p_\beta=0$ and
$z_{\mathrm{ini}}=z_{\mathrm{tp}_2}$), $Nt_1$ and $Nt_2$
increase as:
\begin{eqnarray}
\nonumber      Nt_1 = 0,\,\,\,1,\,\,\,1,\,\,\,2,\,\,\,2,\,\,\,3,\,\,\,3,\,\,\,4...\\
 \nonumber      Nt_2 = 0,\,\,\,0,\,\,\,1,\,\,\,1,\,\,\,2,\,\,\,2,\,\,\,3,\,\,\,3...
\end{eqnarray}

Note the path and limits of integrals in $t$, $\phi$ and $\sigma$ are exactly
the same with those of $p$. We introduce the
following definitions:
\begin{eqnarray}
\begin{aligned}
I_{0\alpha\beta} = &\int^{z_p}_{z_{\mathrm{ini}}}F_{\beta}(z)dz,\; I_{1\alpha\beta}=
\int^{z_{\mathrm{tp}_1}}_{z_p}F_{\beta}(z)dz,\;I_{2\alpha\beta}=\int^{z_p}_{z_{\mathrm{tp}_2}}F_{\beta}(z)dz,\\
&IT_{1\alpha\beta}=\int^{z_{\mathrm{tp}_{1}}}_{z_{_{\mathrm{ini}}}}F_{\beta}(z)dz,\quad
IT_{2\alpha\beta}=\int_{z_{\mathrm{tp}_{2}}}^{z_{_{\mathrm{ini}}}}F_{\beta}(z)dz.
\end{aligned}
\end{eqnarray}
where $\alpha=t$, $\phi$, $\sigma$. Similarly we have
\begin{eqnarray}
I_{1\alpha\beta}=IT_{1\alpha\beta}-I_{0\alpha\beta},\quad
I_{2\alpha\beta}=IT_{2\alpha\beta}+I_{0\alpha\beta}.
\end{eqnarray}
Then the integrals in $t$, $\phi$ and $\sigma$ can be written as
\begin{eqnarray}
\begin{aligned}
\label{integral}
&I_{\alpha\beta}=-\mathrm{sign}(p_\beta) I_{0\alpha\beta}+2Nt_1I_{1\alpha\beta}+2Nt_2I_{2\alpha\beta},\\
&=-[\mathrm{sign}(p_\beta)+2Nt_1-2Nt_2]I_{0\alpha\beta}+2Nt_1IT_{1\alpha\beta}+2Nt_2IT_{2\alpha\beta}.
\end{aligned}
\end{eqnarray}
Finally we have
\begin{eqnarray}
t=I_{t r}+I_{t \theta},\quad \phi=I_{\phi r}+I_{\phi\theta},\quad\sigma = I_{\sigma t}+I_{\sigma \theta}.
\end{eqnarray}

\subsubsection{The computation of elliptic integrals by Carlson's approach}
\label{standartd_forms} In this section we discuss how to compute
the elliptic integrals appeared in $t$, $\phi$ and $\sigma$ by
Carlson's approach. Firstly, we reduce these integrals to the
standard forms. Before the reductions we introduce two notations
$J_k(h)$ and $I_k(h)$ with the following definition:
\begin{eqnarray}
J_k(h) &=& \int^x_y\frac{dt}{(t-h)^k\sqrt{4t^3-g_2t-g_3}},\\
I_k(h) &=&
\int^x_y\frac{dr}{(r-h)^k\sqrt{[(r-u)^2+v^2][(r-w)^2+s^2]}},
\end{eqnarray}
where $k=-2, -1, 0, 1, 2$. From Equations
(\ref{inttheta})-(\ref{intphi}), we have
\begin{eqnarray}
\sigma_\mu &=& t_\mu=
a^2\left[\frac{b_0^2}{16}J_2\left(\frac{b_1}{4}\right)+\frac
{b_0\mu_{\mathrm{tp}_1}}{2}J_1\left(\frac{b_1}{4}\right)+\mu^2_{\mathrm{tp}_1}p\right],\\
\phi_\mu &=& \lambda\left[\frac{
p}{1-\mu_{\mathrm{tp}_1}^2}-\frac{C_-}{2}J_1(z_-)+\frac{C_+}{2}J_1(z_+)\right],
\end{eqnarray}
where
\begin{eqnarray}
  C_{\pm}&=&\frac{b_0}{4(\pm1-\mu_{\mathrm{tp}_1})^2},\\
z_{\pm} &=&
\frac{b_0}{4}\frac{1}{(\pm1-\mu_{\mathrm{tp}_1})}+\frac{b_1}{4}.
\end{eqnarray}
Noting the definition of parameter $p$, we have replaced $J_0$ by
$p$ in the above equations.

\begin{table}
\label{table5}
\begin{center}
    \begin{threeparttable}
\begin{tabular}{c|l}
  \MC{2}{c}{\textbf{Table} 5.\; Standard forms of integrals} \\
  \hline \hline
    \ZZ{-5pt}{15pt} Case &  \MC{1}{c}{$\sigma_r, t_r, \phi_r$}     \\
  \hline
  \begin{minipage}[b]{1em}
             \begin{eqnarray*}
                  1\\
                  \\
             \end{eqnarray*}
   \end{minipage}  &  \begin{minipage}[b]{23em} 
         \begin{eqnarray}
            \sigma_r&=&\frac{b_0^2}{16}J_2\left(\frac{b_1}{4}\right)+
            \frac{b_0r_{\mathrm{tp}_1}}{2}J_1\left(\frac{b_1}{4}\right)+r^2_{\mathrm{tp}_1}p,\\
            t_r &=& \sigma_r+ \nonumber\left[(2+e\varepsilon)(2+r_{\mathrm{tp}_1})-e^2+A_{t+}-A_{t-}\right]p+\\
            &&\frac{(2+e\varepsilon)b_0}{4}J_1\left(\frac{b_1}{4}\right)
            -D_{t+}J_1(z_+)+D_{t-}J_1(z_-),\\
            \phi_r &=& a\left[(A_{\phi+}-A_{\phi-}\right)p-D_{\phi+}J_1(z_+)+D_{\phi-}J_1(z_-)].
         \end{eqnarray}
         \end{minipage}
            \\ \hline
  \begin{minipage}[b]{1em}
             \begin{eqnarray*}
                  2,\;3\\
                  \\
                  \\
             \end{eqnarray*}
   \end{minipage}   &   \begin{minipage}[b][\height]{23em} 
            \begin{eqnarray} 
                \sigma_r &=& \frac{1}{\sqrt{|1-m^2|}}I_{-2}(0),\\
                t_r &=& \nonumber \sigma_r+\frac{1}{\sqrt{|1-m^2|}}[(2+e\varepsilon)I_{-1}(0)+[2(2+e\varepsilon)-e^2]p\\
                && +B_{\mathrm{t+}}I_{1}(r_+)-B_{\mathrm{t-}} I_{1}(r_-)],\\
                \phi_r &=& \frac{a}{\sqrt{|1-m^2|}}\left[B_{\mathrm{\phi+}}I_{1}(r_+)-B_{\mathrm{\phi -}}I_{1}(r_-)\right].
            \end{eqnarray} 
         \end{minipage}
            \\
          \hline
   \begin{minipage}[b]{1em}
             \begin{eqnarray*}
               4,\;5 \\ \\ \\
             \end{eqnarray*}
   \end{minipage}    &   \begin{minipage}[b][\height]{20em} 
            \begin{eqnarray} 
                 \sigma_r &=& \frac{16}{b_0^2}J_{-2}(0)-\frac{8b_1}{b_0^2}J_{-1}(0)+b_1^2J_0(0),\\
                 t_r &=& \sigma_r + \frac{4}{b_0}(2+e\varepsilon)J_{-1}(0)+\left[(2+e\varepsilon)
                         \left(2-\frac{b_1}{b_0}\right)-e^2\right]p\nonumber\\
                 &&+\frac{b_0}{4}\left[B_{\mathrm{t+}}J_1(\tilde{z}_\mathrm{+})-B_{\mathrm{t-}}J_1(\tilde{z}_\mathrm{-})\right],\\
                 \phi_r &=& \frac{ab_0}{4}[B_{\mathrm{\phi +}}J_1(\tilde{z}_\mathrm{+})-B_{\mathrm{\phi -}}J_1(\tilde{z}_\mathrm{-})].
           \end{eqnarray} 
         \end{minipage}
           \\
  \hline \hline
  \MC{2}{c}{\ZZ{-5pt}{15pt}\textbf{Table} 6. \;Definitions of Table 5.} \\
  \hline \hline
    \ZZ{-5pt}{15pt} Case &    \\
   \hline
  \begin{minipage}[b]{1em}
             \begin{eqnarray*}
                  1\\
                  \\
                  \\
                  \\
                  \\
                  \\
             \end{eqnarray*}
   \end{minipage}  &  \begin{minipage}[b]{23em} 
                      \begin{eqnarray*}
                         r_\pm &=& 1\pm\sqrt{1-a^2-e^2},\; z_\pm = \frac{b_0}{4(r_\pm-r_{\mathrm{tp}_1})}+\frac{b_1}{4},\\
                         A_{t\pm} &=& \frac{k_1r_\pm+k_2}{(r_+-r_-)(r_{\mathrm{tp}_1}-r_\pm)},\;
                         D_{t\pm}  =  \frac{(k_1r_\pm+k_2)b_0}{4(r_+-r_-)(r_{\mathrm{tp}_1}-r_\pm)^2},\\
                         k_1 &=& 8-2a\lambda+4(e\varepsilon-e^2)-e^3\varepsilon,\\
                         k_2 &=& e^2(e^2+a\lambda)-2(a^2+e^2)(2+e\varepsilon),\\
                         A_{\phi\pm} &=& \frac{(2-e\varepsilon)r_\pm-(e^2+a\lambda)}{(r_+-r_-)(r_{\mathrm{tp}_1}-r_\pm)},\\
                         D_{\phi\pm} &=& \frac{(2-e\varepsilon)r_\pm-(e^2+a\lambda)}{4(r_+-r_-)(r_{\mathrm{tp}_1}-r_\pm)^2}b_0.
                      \end{eqnarray*}
                   \end{minipage}
               \\ \hline
  \begin{minipage}[b]{1em}
             \begin{eqnarray*}
                  2,\;3
             \end{eqnarray*}
   \end{minipage}   &   \begin{minipage}[b][\height]{23em} 
            \begin{eqnarray*} 
                B_{\mathrm{t\pm}} &=& \frac{k_1 r_{\pm}+k_2}{r_+-r_-},\;
                B_{\mathrm{\phi \pm}} = \frac{(2-e\varepsilon)r_{\pm}-(e^2+\lambda a)}{r_+-r_-}.
            \end{eqnarray*} 
         \end{minipage}
            \\
          \hline
   \begin{minipage}[b]{1em}
             \begin{eqnarray*}
               4,\;5
             \end{eqnarray*}
   \end{minipage}    &   \begin{minipage}[b][\height]{19em} 
            \begin{eqnarray*} 
                 \tilde{z}_\mathrm{\pm} = \frac{b_0}{4}r_{\pm}+\frac{b_1}{4}.
           \end{eqnarray*} 
         \end{minipage}
           \\ \hline\hline
\end{tabular}
\end{threeparttable}
\end{center}
\end{table}

The reduced standard forms of integrals for $r$ are given in Table 5 and 6.
The standard forms for the special cases, such as the equatorial
plane motion and the spherical motion, can be obtained directly and
not given here anymore.

\cite{carlsonQUART,carlson89,carlson91,carlsonDOUBLE}
developed a new approach to compute elliptic integrals
\citep{numrecipes}. He gave new definitions of the standard elliptic
integrals of the first and third kinds
\begin{eqnarray}
R_F(x,y,z)&=&\frac{1}{2}\int^\infty_0\frac{dt}{\sqrt{(t+x)(t+y)(t+z)}},\\
R_J(x,y,z,p)&=&\frac{3}{2}\int^\infty_0\frac{dt}{(t+p)\sqrt{(t+x)(t+y)(t+z)}},
\end{eqnarray}
and the degenerate cases of $R_C(x,y)=R_F(x,y,y)$ and
$R_D(x,y,z)=R_J(x,y,z,z)$. $R_D$ can be regarded as the standard
elliptic integral of the second kind. Carlson denotes the elliptic
integrals by a symbol with the following definition:
\begin{eqnarray}
[p_1,\cdots,p_k]=\int^x_y\prod^k_{i=1}(a_i+b_it)^{p_i/2}dt.
\end{eqnarray}
If $a_i+b_it$ is complex, then its complex conjugate
$\overline{a_i+b_it}$ must exist and guarantee the integral to be
real. And $(a_i+b_it)(\overline{a_i+b_it})=f+gt+ht^2$. Thus
\begin{eqnarray}
\begin{aligned}{}
[p_1,p_1,p_3\cdots,p_k]=\int^x_y(f+gt+&ht^2)^{p_1/2}\\
&\times\prod^k_{j=3}(a_j+b_jt)^{p_j/2}dt.
\end{aligned}
\end{eqnarray}
For a particular elliptic integral, there is an unique formula
to evaluate it. We give a simple example here:
\begin{eqnarray}
[-1,-1,-1,-1]=2R_F(U^2_{12},U^2_{13},U^2_{14}),
\end{eqnarray}
where
\begin{eqnarray*}
  &&U_{ij}=(X_iX_jY_kY_m+Y_iY_jX_kX_m)/(x-y),\\
  &&X_i=\sqrt{a_i+b_ix},\quad Y_i=\sqrt{a_i+b_iy}.
\end{eqnarray*}

The elliptic integrals need to be evaluated in this paper are: $J_1, J_2, I_2,
I_{-1}, I_{-2}$, which can be recast by Carlson's notations. When
equation $4t^3-g_2t-g_3=0$ has three real roots denoted by
$e_1,e_2,e_3$, one has \citep{carlsonQUART}
\begin{eqnarray}
\begin{aligned}
J_k(h)&= s_h\frac{1}{2}\int^x_y\frac{dt}{\sqrt{(t-e_1)(t-e_2)(t-e_3)
(t-h)^{2k}}}\\
&=s_h\frac{1}{2}[-1,-1,-1,-2k],
\end{aligned}
\end{eqnarray}
where $s_h=\mathrm{sign}[(y-h)^k]$. When equation $4t^3-g_2t-g_3=0$
has one pair of complex roots and one real root $e_1$, one has
\citep{carlson91}
\begin{eqnarray}
\begin{aligned}
J_k(h)&= s_h\frac{1}{2}\int^x_y\frac{dt}{\sqrt{(t-e_1)
(t^2+gt+f)(t-h)^{2k}}}\\
&=s_h\frac{1}{2}[-1,-1,-1,-2k].
\end{aligned}
\end{eqnarray}
$I_{k}(h)$ corresponds to the case that equation $R_r(r)=0$ has no
real roots and can be expressed as \citep{carlsonDOUBLE}:
\begin{eqnarray}
\begin{aligned}
 I_{k}(h)&=s_h\int^x_y\frac{dr}{\sqrt{[(r-u)^2+v^2][(r-w)^2+s^2](r-h)^{2k}}},\\
&=s_h[-1,-1,-1,-1,-2k].
\end{aligned}
\end{eqnarray}

Up to now, we have expressed all coordinates and proper time as
functions of parameter $p$ semi-analytically. As discussed in
\cite{yangwang2012}, such treatment is very convenient for massive
particles whose geodesics can be bounded, and the number of times that the particle
meets the turning points can be arbitrary both for $r$ and $\mu$
coordinates. In addition to $p$, one needs to
prescribe the constants of motion. In the next section we discuss
how to get them from the initial four-momentum of a particle.

\section{The constants of motion}
\label{motioncon} As mentioned above, the constants of motion
throughout this paper are defined as
\begin{eqnarray}
\lambda=\frac{L}{E}, q=\frac{Q}{E^2}, m=\frac{\mu_m}{E},
\varepsilon=\frac{\epsilon}{E},
\end{eqnarray}
which can be gotten from the initial four-momentum of a particle
given in a locally nonrotating frame (LNRF) reference. But to
handle more complicated applications, we want to specify the initial
four-momentum in the reference of an assumed emitter, instead of an
LNRF reference directly. However the initial four-momentum is
finally transformed into an LNRF reference by a Lorentz transformation.

Now we introduce the LNRF reference, which is also called as zero
angular momentum observers (ZAMO) \citep{bardeen1972}. The
orthonormal terad is given by
\begin{equation}
\label{simple}\mathbf{e}_{(a)}(\mathrm{LNRF})=e_{{(a)}}^{\nu}\partial_{\nu},
\end{equation}
 where
\begin{eqnarray}
\centering
\label{matrix_e}
e_{{(a)}}^{\nu}=\left(\begin{array}{cccc}
  e^{-\nu}&0 &0 & \omega e^{-\nu}\\
  0       &e^{-\mu_1} &0 & 0\\
  0&0& e^{-\mu_2} &0\\
  0 & 0 & 0 & e^{-\psi}
 \end{array}\right),
\end{eqnarray}
and the dual form of which is
\begin{equation}
\label{simple2}\mathbf{e}^{(a)}(\mathrm{LNRF})=e^{{(a)}}_{\nu}dx^{\nu},
\end{equation}
where
\begin{eqnarray}
 e^{{(a)}}_{\nu}=\left(\begin{array}{cccc}
  e^{\nu}&0 &0 & 0\\
  0       &e^{\mu_1} &0 & 0\\
  0&0& e^{\mu_2} &0\\
  -\omega e^{\psi} & 0 & 0 & e^{\psi}
 \end{array}\right).
\end{eqnarray}
We assume that the particle is shotted by an emitter at the initial
position, where the emitter has coordinate velocities
$\dot{r}=dr/dt$, $\dot{\theta}=d\theta/dt$, and
$\dot{\phi}=\Omega=d\phi/dt$, then its physical velocities
$\upsilon_{r},\upsilon_{\theta},\upsilon_{\phi}$ with respect to the
LNRF fixed at the same point can be written as \citep{bardeen1972}:
\begin{eqnarray}
\label{v_vs_dot_v}\upsilon_{r}=e^{\mu_1-\nu}\dot{r},\quad\upsilon_{\theta}=e^{\mu_2-\nu}\dot{\theta},
\quad\upsilon_{\phi}=e^{\psi-\nu}(\Omega-\omega).
\end{eqnarray}
The orthonormal tetrad of the emitter can be obtained by rotating the
tetrad of the LNRF reference in the local four dimension spacetime.
The rotation is nothing but a Lorentz transformation. We denote the
matrix of the rotation by $\Lambda$, and have
\begin{eqnarray}
\label{tetrad_obs_LNRF}\mathbf{e}_{(a)}(\mathrm{em})=\Lambda^{(b)}_{(a)}
\mathbf{e}_{(b)}(\mathrm{LNRF}),
\end{eqnarray}
where \citep{mtw}
\begin{eqnarray}
\label{matrix2}\Lambda_{(a)}^{(b)}= \left(\begin{array}{ccc}
\gamma & \gamma\upsilon_r & \gamma\upsilon_\theta \\
\gamma\upsilon_r & 1+\gamma^2\upsilon^2_r/(1+\gamma) & \gamma^2\upsilon_r\upsilon_\theta/(1+\gamma)\\
\gamma\upsilon_\theta & \gamma^2\upsilon_\theta\upsilon_r/(1+\gamma) & 1+\gamma^2\upsilon_\theta^2/(1+\gamma) \\
\gamma\upsilon_\phi & \gamma^2\upsilon_\phi\upsilon_r/(1+\gamma) &
\gamma^2\upsilon_\phi\upsilon_\theta/(1+\gamma)
\end{array}\right. \nonumber
\\ \left.\begin{array}{cc}
  &\gamma\upsilon_\phi\\
  &\gamma^2\upsilon_r\upsilon_\phi/(1+\gamma)\\
  &\gamma^2\upsilon_\theta\upsilon_\phi/(1+\gamma)\\
  &1+\gamma^2\upsilon_\phi^2/(1+\gamma)
\end{array}\right),
\end{eqnarray}
where
$\gamma=[1-(\upsilon_r^2+\upsilon_{\theta}^2+\upsilon_{\phi}^2)]^{-1/2}$,
and the covariant tetrad of the emitter is
\begin{equation}
\mathbf{e}^{(a)}(\mathrm{em})=(\Lambda^{-1})^{(a)}_{(b)}
\mathbf{e}^{(b)}(\mathrm{LNRF}),
\end{equation}
where
\begin{eqnarray}
\label{matrix1}(\Lambda^{-1})^{(b)}_{(a)}= \left(\begin{array}{ccc}
\gamma & -\gamma\upsilon_r & -\gamma\upsilon_\theta \\
-\gamma\upsilon_r & 1+\gamma^2\upsilon^2_r/(1+\gamma) & \gamma^2\upsilon_r\upsilon_\theta/(1+\gamma) \\
-\gamma\upsilon_\theta & \gamma^2\upsilon_\theta\upsilon_r/(1+\gamma) & 1+\gamma^2\upsilon_\theta^2/(1+\gamma) \\
-\gamma\upsilon_\phi & \gamma^2\upsilon_\phi\upsilon_r/(1+\gamma) &
\gamma^2\upsilon_\phi\upsilon_\theta/(1+\gamma)
\end{array}\right.\nonumber \\
\left.\begin{array}{cc}
 & -\gamma\upsilon_\phi\\
 & \gamma^2\upsilon_r\upsilon_\phi/(1+\gamma)\\
 & \gamma^2\upsilon_\theta\upsilon_\phi/(1+\gamma)\\
 & 1+\gamma^2\upsilon_\phi^2/(1+\gamma)
\end{array}\right).
\end{eqnarray}
Equivalently, one has
\begin{eqnarray}
&&\label{lnrfcon}\mathbf{e}_{(a)}(\mathrm{LNRF})=(\Lambda^{-1})^{(b)}_{(a)}
\mathbf{e}_{(b)}(\mathrm{em}), \\
&&\mathbf{e}^{(a)}(\mathrm{LNRF})=(\Lambda)^{(a)}_{(b)}
\mathbf{e}^{(b)}(\mathrm{em}).
\end{eqnarray}
In the rest frame of the emitter, the components of four-momentum of
the particle are denoted by $p'_{(a)}$, which can be regarded as the
projections of the momentum $\mathbf{p}$ on the corresponding basis
vectors, i.e.,
\begin{equation}
\label{comobser}p'_{(a)}=\mathbf{p}\cdot
\mathbf{e}_{(a)}(\mathrm{em}).
\end{equation}
Multiplying both sides of
Equation (\ref{lnrfcon}) by $\mathbf{p}$, we get
\begin{eqnarray}
\label{p_and_pprime1}e^{\nu}_{(a)}p_\nu
=(\Lambda^{-1})_{(a)}^{(b)}p'_{(b)},
\end{eqnarray}
Assuming that the physical velocities of the particle with respect
to the emitter are:
$\upsilon_r',\,\,\upsilon_\theta',\,\,\upsilon_\phi'$, we have
\begin{eqnarray}
p'_{(a)}=\gamma'\mu_m(-1,\upsilon_r',\,\,\upsilon_\theta',\,\,\upsilon_\phi'),
\end{eqnarray}
where $\gamma' =[1-(\upsilon_r'^2+\upsilon_\theta'^2+\upsilon_\phi'^2)]^{-1/2}$ is
the Lorentz factor. Equation (\ref{p_and_pprime1}) can be expanded
explicitly by using Equations (\ref{fourmomentum}) and
(\ref{matrix_e}):
\begin{eqnarray}
\label{energy_t}Ee^{-\nu}(1-\lambda\omega)&=&\gamma'\mu_m k_{(t)},\\
\label{energy_r}Ee^{-\mu_1}s_r\frac{\sqrt{R_r}}{\Delta}&=&\gamma'\mu_m k_{(r)},\\
\label{energy_theta}Ee^{-\mu_2}s_{\theta}\sqrt{\Theta_\theta}&=&\gamma'\mu_m k_{(\theta)},\\
\label{energy_phi}E\lambda e^{-\psi}&=&\gamma'\mu_m k_{(\phi)},
\end{eqnarray}
where
\begin{eqnarray}
\label{energy_t1}k_{(t)}&=&\gamma(1+\mathbf{v}\cdot\mathbf{v'}),\\
\label{energy_r1}k_{(r)}&=&\gamma\upsilon_r+\upsilon_r'+\frac{\gamma^2\upsilon_r}{1+\gamma}
\mathbf{v}\cdot\mathbf{v'},\\
\label{energy_theta1}k_{(\theta)}&=&\gamma\upsilon_\theta+\upsilon_\theta'+\frac{\gamma^2\upsilon_\theta}{1+\gamma}
\mathbf{v}\cdot\mathbf{v'},\\
\label{energy_phi1}k_{(\phi)}&=&\gamma\upsilon_\phi+\upsilon_\phi'+\frac{\gamma^2\upsilon_\phi}{1+\gamma}
\mathbf{v}\cdot\mathbf{v'},\\
\mathbf{v}\cdot\mathbf{v'}&=&\upsilon_r\upsilon_r'+\upsilon_\theta\upsilon_\theta'+\upsilon_\phi\upsilon_\phi'.
\end{eqnarray}

\begin{figure}
\centering
\includegraphics[width=0.4\textwidth,angle=0]{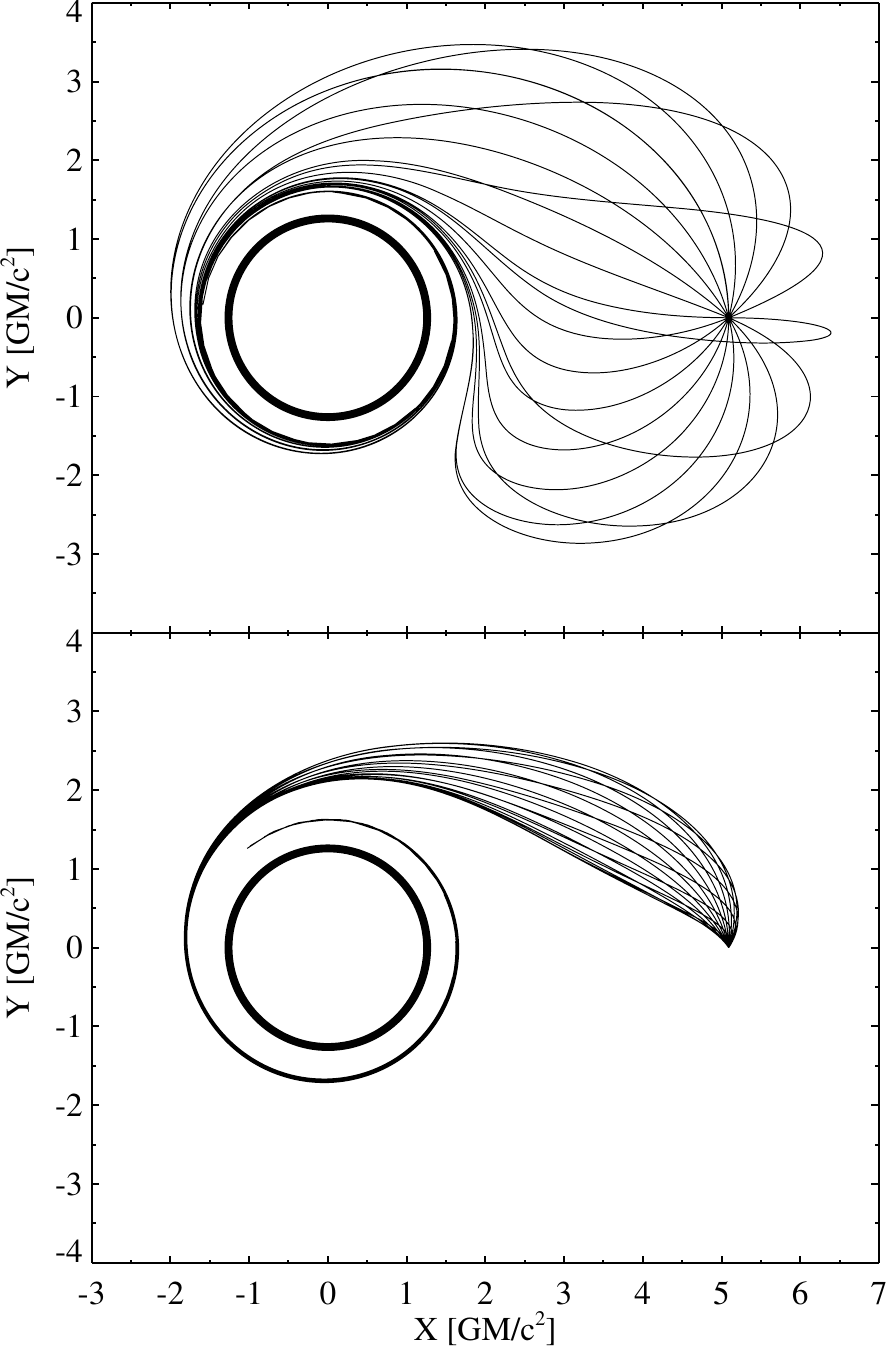}
\caption{\label{posi_neg} This figure shows the trajectoris of a set
of particles, which are confined in the equatorial plane and do
plane motions. The spin and electric charge of the black hole are
0.96 and 0.1 respectively. The electric charges of particles are
$\varepsilon_{+}$ (top panel) and $\varepsilon_{-}$ (bottom panel). The
initial physical speed of the particles is isotropical and equal to
0.35 with respect to the LNRF. The initial coordinates of the
particles are $r=5$ $r_\mathrm{g}$, $\theta=90^\circ$, and $\phi=0$.
A circle at the center represents the event horizon.}
\end{figure}

Solve Equations (\ref{energy_t}) and (\ref{energy_phi}) for
$\lambda$, we get
\begin{eqnarray}
\lambda = \frac{k_{(\phi)}\sin\theta}{k_{(t)}\sqrt{\Delta}
\Sigma/A+k_{(\phi)}\omega\sin\theta}.
\end{eqnarray}
With $\lambda$, from Equation (\ref{energy_t}), we get
\begin{eqnarray}
m = \frac{e^{-\nu}}{\gamma'k_{(t)}}(1-\lambda \omega).
\end{eqnarray}
With $\lambda$ and $m$, from Equation (\ref{energy_theta}), we get
\begin{eqnarray}
\begin{aligned}
q =
\left[\left(\frac{k_{(\phi)}}{k_{(t)}\sqrt{\Delta}\Sigma/A+k_{(\phi)}\omega\sin\theta}\right)^2+a^2(m^2-1)\right]
\cos^2\theta\\
+m^2\Sigma\gamma'^2k^2_{(\theta)}.
\end{aligned}
\end{eqnarray}
With $\lambda$, $m$, and $q$, from Equation (\ref{energy_r}), we get
a quadratic equation for $\varepsilon$,
\begin{eqnarray}
a_1\varepsilon^2+b_1\varepsilon+c_1=0,
\end{eqnarray}
where
\begin{eqnarray}
a_1&=&e^2r^2, \\
b_1&=&2er(r^2+a^2-a\lambda),\\
\nonumber  c_1&=&(1-m^2)r^4+2m^2r^3-[q+\lambda^2+a^2(m^2-1)+e^2m^2]r^2\\
\nonumber &&+2[q+(a-\lambda)^2]r-e^2(a-\lambda)^2-(a^2+e^2)q\\
&&-\Delta\Sigma(m\gamma'k_{(r)})^2.
\end{eqnarray}
We then get two values,
\begin{eqnarray}
\varepsilon_{\pm}=\frac{-b_1\pm\sqrt{b_1^2-4a_1c_1}}{2a_1},
\end{eqnarray}
where $\varepsilon_+$ represents positive electric charge, and
$\varepsilon_-$ represents negative electric charge. In Figure
\ref{posi_neg}, we plot a set of geodesic orbits of particles
emitted isotropically in an LNRF reference. These particles are
confined in the equatorial plane. The particles have
$\varepsilon_{+}$ in the top panel, and $\varepsilon_{-}$ in the
bottom panel.

\section{A brief introduction to the code}
\label{codeintroduction} According to the discussions above, we have
developed a new public code for computing null and time-like
geodesics in a K-N spacetime \footnote{The source FORTRAN
code can be download on our Web site
\url{http://www1.ynao.ac.cn/~yangxl/yxl.html}}. We name the code $ynogkm$, which is written in
fortran 95, and the object oriented method is used. The code
consists of several independent modules, in which each one completes
a special goal. The most important two modules are $ellfunction$
and $blcoordinates$. The former one contains the supporting functions
and subroutines computing elliptic integrals by Carlson's approach.
The latter one contains the functions and routines computing the B-L
coordinate functions: $r(p)$, $\mu(p)$, $\phi(p)$, and $t(p)$, as
well as proper time function $\sigma(p)$. In $blcoordinates$, we
provide a subroutine named $ynogkm$ to compute all coordinates and
proper times simultaneously for a given $p$. We also provide two
functions named $radiusm$ and $mucosm$ to compute $r(p)$ and
$\mu(p)$ respectively. In an axis-symmetry case, one only needs to
compute $r$ and $\mu$.

Before calling these functions and subroutines to compute the B-L
coordinates, one needs to provide the constants of motion, namely,
$\lambda$, $q$, $m$, and $\varepsilon$. As discussed in the above
section, we have provided a set of formulae to compute these
constants from $\upsilon_{i}'$, which are the physical velocities of
the particle with respect to an assumed emitter, who has also
physical velocities $\upsilon_{i}$ with respect to an LNRF reference.
According to these formulae, we provide a subroutine named
$lambdaqm$ to calculate $\lambda$, $q$, $m$, $\varepsilon$, and
$k_{(a)}$ defined by Equations (\ref{energy_t1})-(\ref{energy_phi1}).
Except for a factor $\gamma'\mu_m$,
$k_{(a)}$ actually are exactly equal to the initial four-momentum of the text
particle given in an LNRF. Thus $k_{(a)}$ can be used to determine the
signs in front of $\Pi_r$ or $\Pi_\mu$. The other initial parameters
need to be specified are included: (1) the initial coordinates of the particle,
$r_{\rms{ini}}$, $\theta_{\rms{ini}}$, $\phi_{\rms{ini}}$ and $t_{\rms{ini}}$.
The latter two ones are usually set to be zero; (2) the
physical velocities of the assumed emitter with respect to an LNRF,
$\upsilon_{r}$, $\upsilon_{\theta}$, and $\upsilon_{\phi}$; (3) the
physical velocities of the particle with respect to the assumed emitter,
$\upsilon_{r}'$, $\upsilon_{\theta}'$, and $\upsilon_{\phi}'$; (4)
the spin parameter $a$ and the electric charge $e$ of the black
hole. With a given $p$ and those initial parameters, one can do the
calculations directly without giving the number of times that the particle
meets the two turning points, namely $Nt_1$ and $Nt_2$.

In our code the parameter $p$ is an independent variable, which is
always positive and monotonously increasing along a particular geodesic. When
the geodesic is unbounded, it has a termination, either at infinity
or the event horizon. The value of $p$ corresponding to the
termination is a finite number, denoted by $p_{\mathrm{max}}$. We
provide a subroutine named $ptotal$ to calculate this number.
Apparently, when a $p$ given by the user is bigger than
$p_{\mathrm{max}}$, it has no meaning and the code resets it to be
$p_{\mathrm{max}}$ mandatorily. When a geodesic is bounded, its
termination does not exist at all and $p$ can take any positive
value.

For a more detailed introduction, one can see the
README\footnote{\url{http://www1.ynao.ac.cn/~yangxl/ynogkm/readme.pdf}} file. In
the next section, we give the results of our code for toy problems.

\section{Applications for toy problems}
\label{protest} To show the utility of our code, we apply it to
toy problems. The results for five such examples are illustrated in this section.

\begin{figure*}
\begin{center}
\includegraphics[width=0.9\textwidth,angle=0]{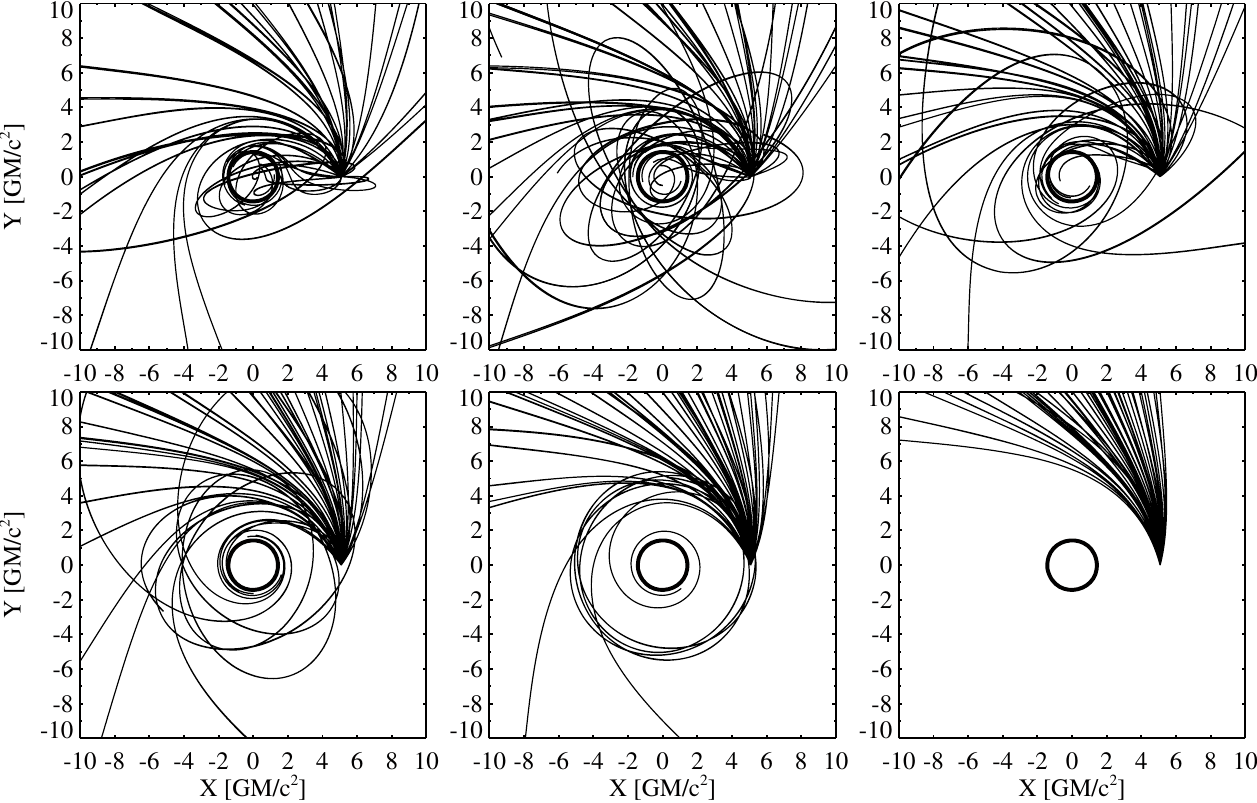}
\caption{\label{rays} A set of geodesics of massive particles orbit
around a black hole with $a$=0.9. The physical velocities of the
emitter with respect to the LNRF are $\upsilon_r=0$,
$\upsilon_\theta=0$, and $\upsilon_\phi=0.4$, $\upsilon_\phi=0.5$,
$\upsilon_\phi=0.6$, $\upsilon_\phi=0.7$, $\upsilon_\phi=0.8$,
$\upsilon_\phi=0.9$ for panels from left to right, top to bottom.
The velocities of these particles are specified isotropically in the
rest frame of the emitter and $\upsilon_{\mathrm{pt}}'=0.6$.}
\end{center}
\end{figure*}

\subsection{Geodesic orbits of massive particles}
The most important application of the code is to compute the
geodesics of massive particles in a K-N spacetime. As the
first application, we use the code to compute the orbits of a set of test
particles that are emitted isotropically in the local rest frame
of an assumed emitter. The particles have a constant speed
$\upsilon_{\mathrm{pt}}'$ but different directions in the local reference, and
$\upsilon_{\mathrm{pt}}'=0.6$. The orientation of the
velocity is described by $\vartheta$ and $\varphi$, thus the components
of the velocity under the reference of the emitter are

\begin{eqnarray}
\upsilon_r' &=& \upsilon_{\mathrm{pt}}'\sin\vartheta\cos\varphi,\\
\upsilon_\theta' &=& \upsilon_{\mathrm{pt}}'\sin\vartheta\sin\varphi,\\
\upsilon_\phi' &=& \upsilon_{\mathrm{pt}}'\cos\vartheta.
\end{eqnarray}
The physical velocities of the emitter with respect to the LNRF
reference are $\upsilon_r=0$, $\upsilon_\theta=0$, in which only the
$\phi$ component is not zero and takes different values. We
demonstrate the results in Figure \ref{rays}. It is shown that as
the speed of emitter increases, more particles become unbounded, and
the beaming effect becomes more significant.

\subsection{The orbits of spherical motion}
The circular orbits in the Kerr spacetime has significant
applications in the standard geometrically thin accretion disk
systems. A particle in the accretion flow loses its angular momentum
by viscosity and moves inward slowly. Its angular velocity is far
greater than its radial velocity. Thus the particle moves in a
circular orbit is a good approximation. The inner radius of the disk
is usually located at the ISCO. Based on this assumption, one can
measure the black hole spin by fitting the line profiles or the
continuous spectra. Actually the circular orbits, in which the
particle is confined in the equatorial plane of the black hole, can
be regarded as a special case of the spherical orbit. The radial
velocity and acceleration of the particle in a spherical orbit are
vanished, leading to two conditions: $dr/d\tau=0$ and
$d^2r/d\tau^2=0$. Using equation (\ref{defr}), these conditions
reduce to \citep{bardeen1972,wilkins1972}:
\begin{eqnarray}
R_r=0,\quad \frac{dR_r}{dr}=0.
\end{eqnarray}
We use $\theta_*$ to denote the coordinate of one of the $\theta$
turning points, therefore we have $\Theta_\theta(\theta_*)$=0. Using
the same strategy discussed in \cite{shakura1987}, we can get the
angular velocity of the particle at $\theta_{*}$
\begin{eqnarray}
\Omega^* = \frac{\sqrt{P}}{\sin\theta_*
(\pm\Sigma\sqrt{r}+a\sin\theta_*\sqrt{P}) },
\end{eqnarray}
where $P=M(r^2-a^2\cos^2\theta_*)-e^2r$, and the constants of
motion:
\begin{eqnarray}
\label{evm}\frac{E}{\mu_m} &=& \frac{\pm a
\sin\theta_*\sqrt{P}+(\Delta-a^2\sin^2\theta_*)\sqrt{r}}{\sqrt{\Sigma}
\sqrt{-P+(\Delta-a^2\sin^2\theta_*)r\pm2a\sin\theta_*\sqrt{rP}}}, \\
\label{lvm}\frac{L}{\mu_m}&=&
\frac{\sin\theta_*[\pm(r^2+a^2)\sqrt{P}-\sqrt{r}a\sin\theta_*(2Mr-e^2)]
}{\sqrt{\Sigma}\sqrt{-P+(\Delta-a^2\sin^2\theta_*)r\pm2a\sin\theta_*\sqrt{rP}}},\\
\nonumber\label{qvm}
\frac{Q}{\mu_m^2}&=&\cos^2\theta_*\left[a^2\left(1-\frac{E^2}{\mu_m^2}\right)+
\frac{1}{\sin^2\theta_*}\frac{L^2}{\mu_m^2}\right],\\
&=&\frac{r\cos^2\theta_*\left[A_Q\mp2a\sin\theta_*
(2Mr-e^2)r\sqrt{rP}\right]}{ \Sigma[
-P+(\Delta-a^2\sin^2\theta_*)r\pm2a\sin\theta_*\sqrt{rP}]},
\end{eqnarray}
where
\begin{eqnarray}
\begin{aligned}
 A_Q = (r^2+a^2)^2&(Mr-e^2)+a^2\{[Mr(r^2-a^2)\\
 &+e^2a^2]\sin^2\theta_*-(2Mr-e^2)^2\cos^2\theta_*\}.
\end{aligned}
\end{eqnarray}
In these formulae, the upper sign refers to the prograde orbits
(i.e., corotating with L$>$0), while the lower sign refers to
retrograde orbits (counter rotating with L$<$0).

\begin{figure}
\begin{center}
\includegraphics[width=0.4\textwidth,angle=0]{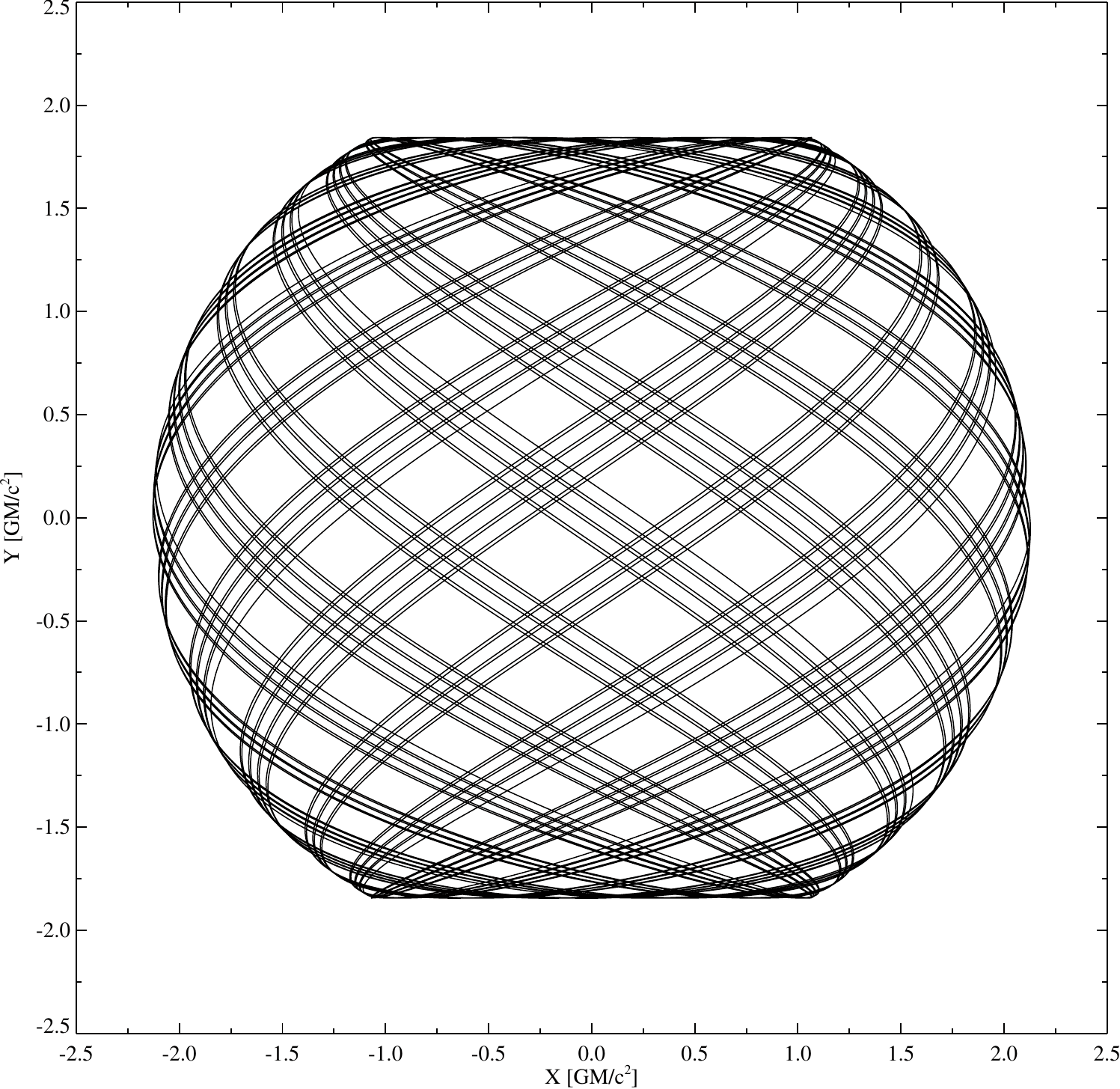}
\caption{\label{sphericalmotion1} The orbit of a particle in a
spherical motion. The parameters are: the black hole spin $a$=0.998,
the radius of the orbit r=2 $r_\mathrm{g}$, the turning point
$\theta_*=30^\circ$, the inclination angle of the observer
$\theta_{\mathrm{obs}}=90^\circ$.}
\end{center}
\end{figure}

In Figure \ref{sphericalmotion1}, we plot the orbit of a particle in
a spherical motion. Given the parameters: $a, e, r,$ and $\theta_*$,
from Equations (\ref{evm})-(\ref{qvm}), we can get the constants of
motion: $\lambda, q, m$, with which from equation $R_r=0$ (or
$dR_r/dr=0$), we can get the final constant $\varepsilon$. For
simplicity, we let both $e$ and $\varepsilon$ to be zero in this
figure. Comparing with circular motion, the most significant effect
of spherical motion is the precession of the orbit.

Correspondingly, the spherical motion has also three kinds of
marginal orbits, which are:\\
(1). Photon orbit $r_{\mathrm{ph}}$, which is the innermost boundary
of the spherical orbits for particles, it occurs when the
denominator of Equations (\ref{evm}), (\ref{lvm}), and (\ref{qvm})
vanishes, i.e.,
\begin{eqnarray}
-P+(\Delta-a^2\sin^2\theta_*)r\pm2a\sin\theta_*\sqrt{rP}=0.
\end{eqnarray}
(2). Marginally bound spherical orbit $r_{\mathrm{mb}}$, which occurs when $E/\mu_m=1$.\\
(3). Inner most marginally stable spherical orbit $r_{\mathrm{ms}}$ (ISSO).
The stable condition requires that $d^2R_r/dr^2\leq0$, which yields
the equivalent condition,
\begin{eqnarray}
1-\left.\left(\frac{E}{\mu_m}\right)^2\right|_{e=0}\geq
\frac{2Mr(r^2-3a^2\cos^2\theta_*)}{3r^4-a^4\cos^4\theta_*-6r^2a^2\cos^2\theta_*},
\end{eqnarray}
or $r\geq r_{\rms ms}$. For simplicity we have let $e$ to be zero in the
above equation. If $\theta_*=\pi/2$, namely the circular orbits,
this condition reduces to the same form of Equation (2.20) of
\cite{bardeen1972}.

\begin{figure}
\begin{center}
\includegraphics[width=0.4\textwidth,angle=0]{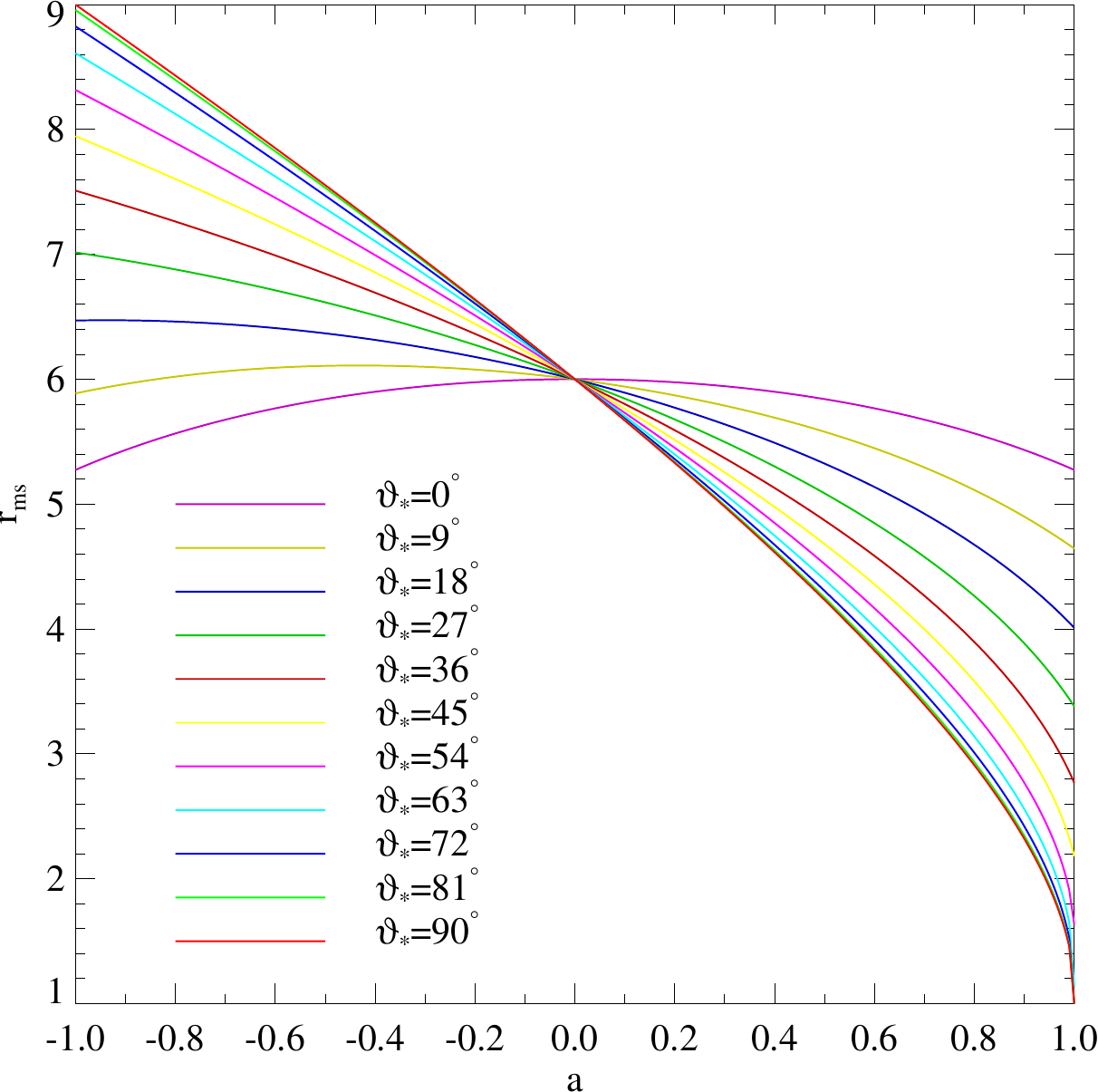}
\caption{\label{rms} The radii of the innermost stable orbits for
the spherical motion around a Kerr black hole, as functions of the
specific angular momentum $a$ of the black hole.}
\end{center}
\end{figure}

In our code, we provide three functions named $r\_\mathrm{ms},
r\_\mathrm{mb},$ and $r\_\mathrm{ph}$ to compute the radii of these
orbits with given $a, \theta_*, e$. In Figure
\ref{rms}, we plot the radii of inner most stable spherical orbits
as functions of $a$ for various $\theta_*$. For simplicity we
let also $e=0$ in this Figures. One can see that as $\theta_*$
increases, the radii become larger for $a>0$ and smaller for $a<0$.
For $a=0$, the radii keeps unchanged. When $\theta_*=0^\circ$, the
curve becomes symmetry for $a>0$ and $a<0$. The similar properties
can be obtained for the radii of photon orbits and marginally bound
orbits.

\subsection{Orbits inside $r_{\mathrm{ms}}$}
\begin{figure}
\begin{center}
\includegraphics[width=0.4\textwidth,angle=0]{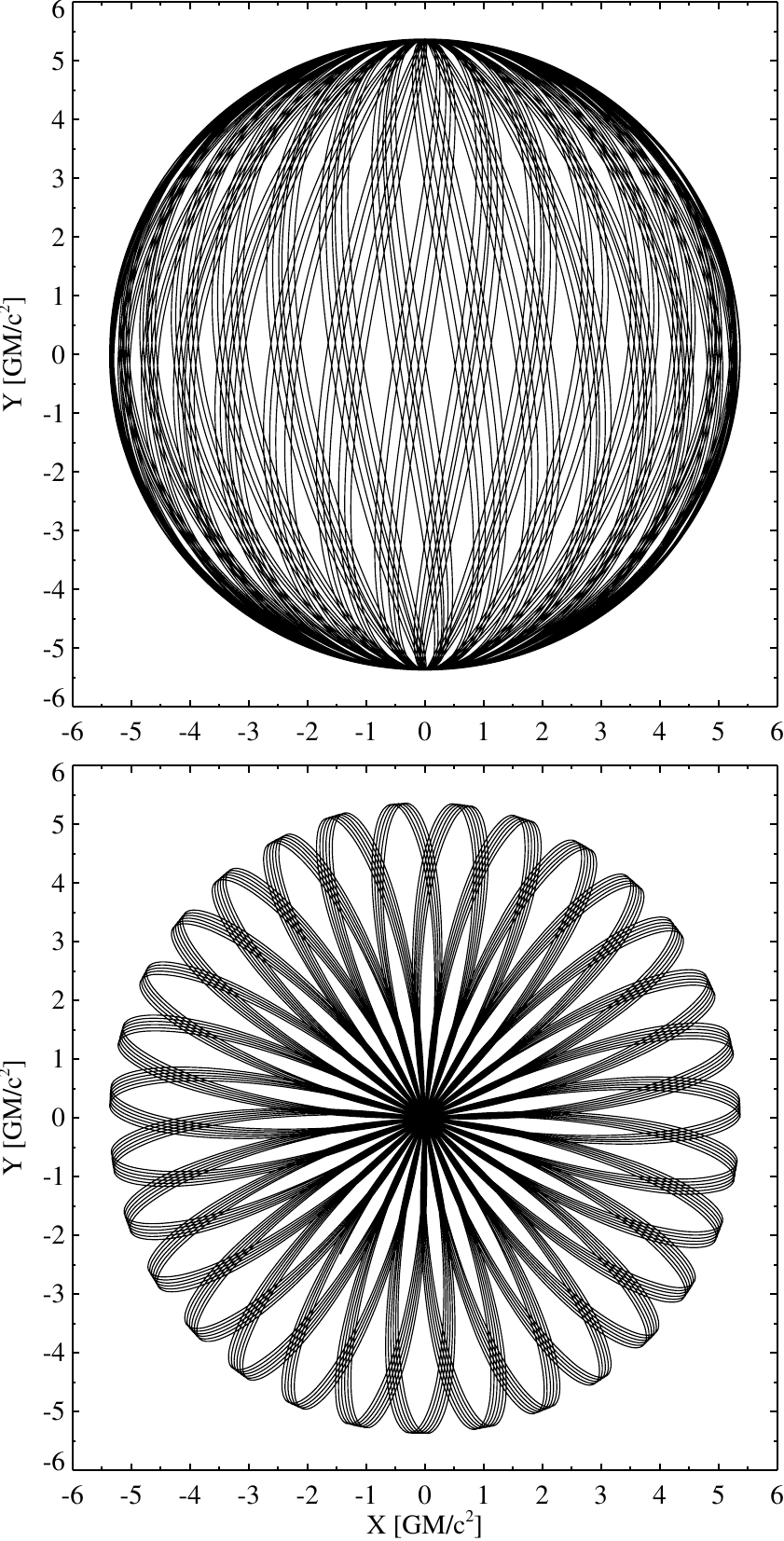}
\caption{\label{sphericalmotion} The orbit of a particle moving
inside the ISSO with the constants of motion of the ISSO. The balck
hole spin $a$=0.998 and the turning point $\theta_*=0^\circ$. The
radius of ISSO is $r=5.2781$ $r_\mathrm{g}$. The inclination angle
$\theta_{\mathrm{obs}}=90^\circ$ for top panel and
$\theta_{\mathrm{obs}}=0^\circ$ for bottom panel.}
\end{center}
\end{figure}

The region inside ISSO usually called as the plug region, in which a
particle moves along geodesics with constants of motion of the
marginally stable spherical geodesic \citep{cunningham1975} when its
initial radial perturbation velocity $\upsilon_r=0$. With the
results presented in the above section, we get the constants of
motion for the marginally stable spherical orbits
\begin{eqnarray}
\left(\frac{E}{\mu_m}\right)^2&=&1-\frac{2Mr_{\mathrm{ms}}(r_{\mathrm{ms}}^2-
3a^2\cos^2\theta_{*})}{D_{\mathrm{ms}}},\\
\left(\frac{L}{\mu_m}\right)^2&=&2Mr_{\mathrm{ms}}\sin^2\theta_{*}[3r_{\mathrm{ms}}^4-r_{\mathrm{ms}}^2a^2(1+
\cos^2\theta_{*})\nonumber \\
&&\qquad+3a^4\cos^2\theta_{*}]/D_{\mathrm{ms}},\\
\frac{Q}{\mu^2_m}&=&\frac{2Mr_{\mathrm{ms}}^3\cos^2\theta_{*}(3r_{\mathrm{ms}}^2-
a^2\cos^2\theta_{*})}{D_{\mathrm{ms}}},\\
D_{\mathrm{ms}} &=&
3r_{\mathrm{ms}}^4-a^4\cos^4\theta_{*}-6r_{\mathrm{ms}}^2a^2\cos^2\theta_{*},
\end{eqnarray}
and
\begin{eqnarray}
\begin{aligned}
R_r=2M(r_{\rms{ms}}-r)^3[rr_{\rms{ms}}^3-3a^2&r_{\rms{ms}}
\cos^2\theta_{*}(r_{\rms{ms}}+r) \\
&+a^4\cos^4\theta_{*}]/D_{\rms{ms}},
\end{aligned}\\
\begin{aligned}
\Theta_\mu=2Mr_{\rms{ms}}(\cos^2\theta_{*}-&\mu^2)
[(3a^4\cos^2\theta_{*}-r_{\rms{ms}}^2a^2)\mu^2   \\
&-r_{\rms{ms}}^2a^2\cos^2\theta_{*}+3r_{\rms{ms}}^4]/D_{\rms{ms}}.
\end{aligned}
\end{eqnarray}
From the above two equations, we know that both $r_{\mathrm{ms}}$
and $\theta_{*}$ are the turning points. With these expressions, we
can get the constants of motion immediately to compute the geodesic
orbits inside $r_{\mathrm{ms}}$. In Figure \ref{sphericalmotion}, we
plot such an orbit. We take $a=0.998$ and $\theta_{*}=0$, namely the
particle goes through the spin axis of the black hole. Using the
function $r\_\mathrm{ms}(a,\theta_*)$ in our code, we get
$r_{\mathrm{ms}}=5.2781$ $r_\mathrm{g}$. From this figure, one can
see that the orbit is almost the same with a spherical motion,
because the radial velocity is much smaller than the poloidal and
azimuthal velocities.

\subsection{The accretion flow of disk}
Now we use our code to construct a toy model for mimicking accretion
flows of skewed geometrically thin disks. The flows are composed by
non-interacting particles, which fall freely into the black hole
along the geodesic trajectories. It implies that we make a ballistic
treatment to the fluid flow, and the dynamics and the structure of
the disk are uniquely determined by the gravitational field of the
black hole. It is also convenient to regard the accretion flow as a
collection of test particles with same mass. The boundary conditions
of the disk are assumed to be a ring at $r=r_0$, from which the test
particles are continuously injected. The plane of the ring has an
inclination angle $\beta$ with respect to the spin axis.

To describe the initial conditions of the test particles, namely the
velocities, an orthonormal tetrad is established on the ring. We
choose three spacial basis vectors of the tetrad as :
$\mathbf{e}_x,\mathbf{e}_y,\mathbf{e}_z$, where $\mathbf{e}_x$ is
the tangent vector of the ring, $\mathbf{e}_y$ is aligned along the
radial direction and pointed inward, $\mathbf{e}_z$ is the normal
vector of the plane of the ring, and these vectors satisfy
right-hand rule. In the local rest frame of the tetrad, the physical
velocities of the particle are $\upsilon'_x, \upsilon'_y,
\upsilon'_z$. According to the discussions in Section
\ref{motioncon}, in order to compute the constants of motion from
these velocities, we need to transform them into the LNRF reference
for getting $\upsilon_r, \upsilon_\theta, \upsilon_\phi$.

\begin{figure}
\centering
\includegraphics[width=0.4\textwidth,angle=0]{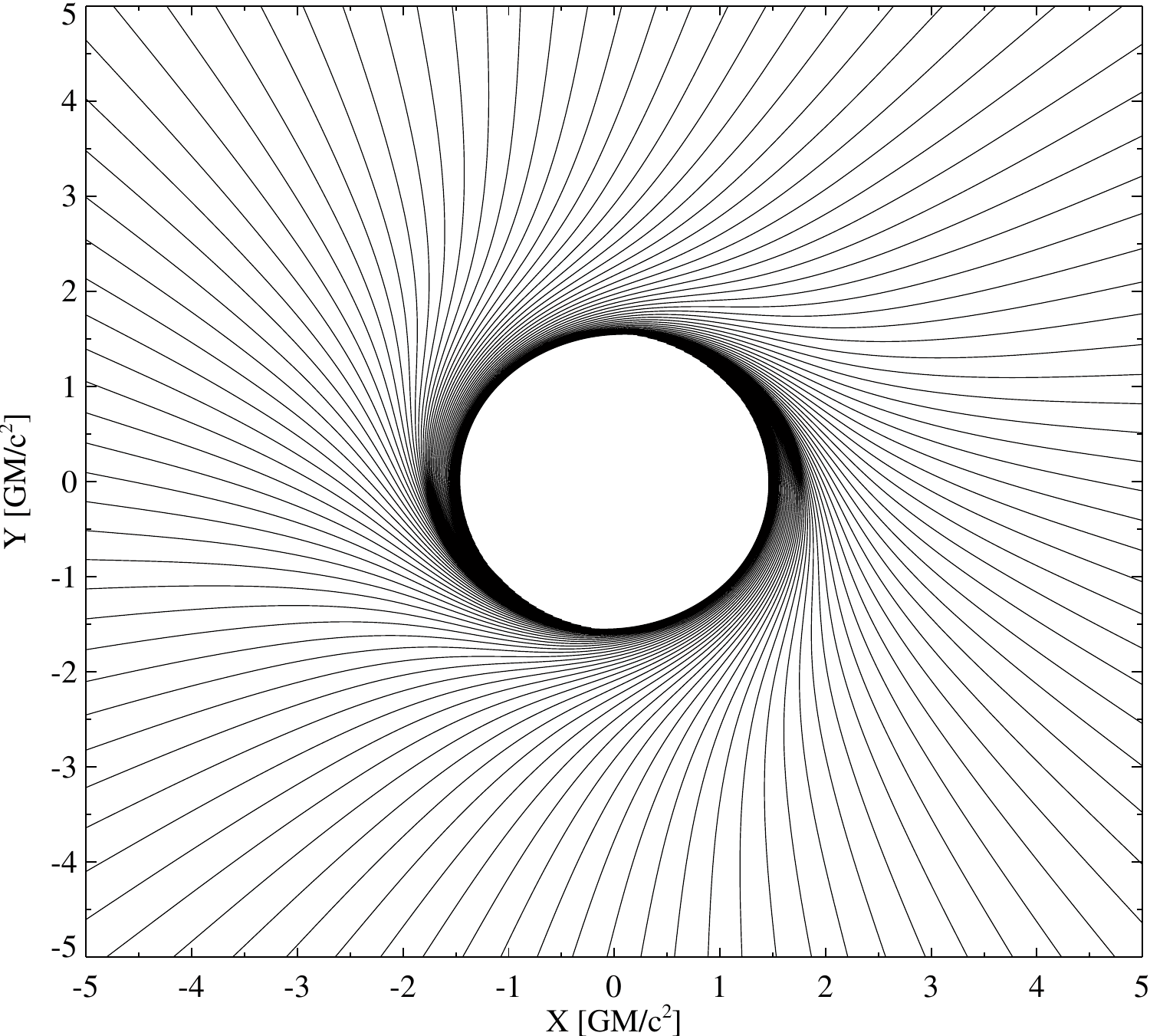}
\caption{\label{orbitsdisk} The orbits of equal mass test particle
flow inward to a black hole along geodesic trajectories, which form
a smooth and curved surface. The black hole spin $a=0.96$, the
initial tiled angle of the disk is $\beta=30^\circ$. The initial
physical velocities of the particles under the tetrad are:
$\upsilon_x'=0.01$, $\upsilon_y'=0.5$, $\upsilon_z'=-0.01\cos\phi$.
The radius of the ring is 20 $r_\mathrm{g}$.}
\end{figure}

We denote the transformation matrix by $T(\theta, \phi, \beta)$. The
explcit expression of $T$ is presented in Appendix \ref{matrixT}.
Hence we have $\upsilon_{i}=T_i^j\upsilon_j'$. Since the tetrad is
attached on the ring, $\theta$ and $\phi$ are not independent
variables, actually they satisfy the following equation
\begin{eqnarray}
\centering \sin\theta =
\frac{\cos\beta}{\sqrt{1-\sin\phi^2\sin^2\beta}}.
\end{eqnarray}
Therefore we have $\upsilon_{i}=T_i^j(\phi,\beta)\upsilon_j'$.
Taking $\phi$ as an independent variable that varies from 0 to
$2\pi$, we can set the initial conditions for all test particles. In
Figure \ref{orbitsdisk}, we plot such a set of geodesic orbits of
test particles with same mass. The physical velocities are
$\upsilon_x'=0.01$, $\upsilon_y'=0.5$, and $\upsilon_z'=-0.01$ respectively.
From the figure, one can see that all trajectories form a smooth but
curved surface.

Using the ray-tracing approach \citep{Luminet1979}, we can image the curved surface. In
\cite{yangwang2012}, we have presented a new public code named ynogk
to compute the null geodesics in a Kerr spacetime and a more general
method to image a target object. The method requires one to provide
the function describing the surface, i.e., $F(r,\theta,\phi)=0$, or
$F(x,y,z)=0$. For this curved surface formed by geodesic orbits of
test particles, we can not write out its explicit form, and only
use the interpolation approach. To approximate the surface, we take
N particles with N geodesic orbits. We take M points in each orbit,
and totally get N$\times$M points. We can get the coordinates of
each point easily and write them as: $z_i=z_i(x_i,y_j)$. Using the
interpolation approaches provided in \cite{numrecipes}, we can get a
approximation function $z=z(x,y)$ that describes the surface.

\begin{figure}
\centering
\includegraphics[width=0.42\textwidth,angle=0]{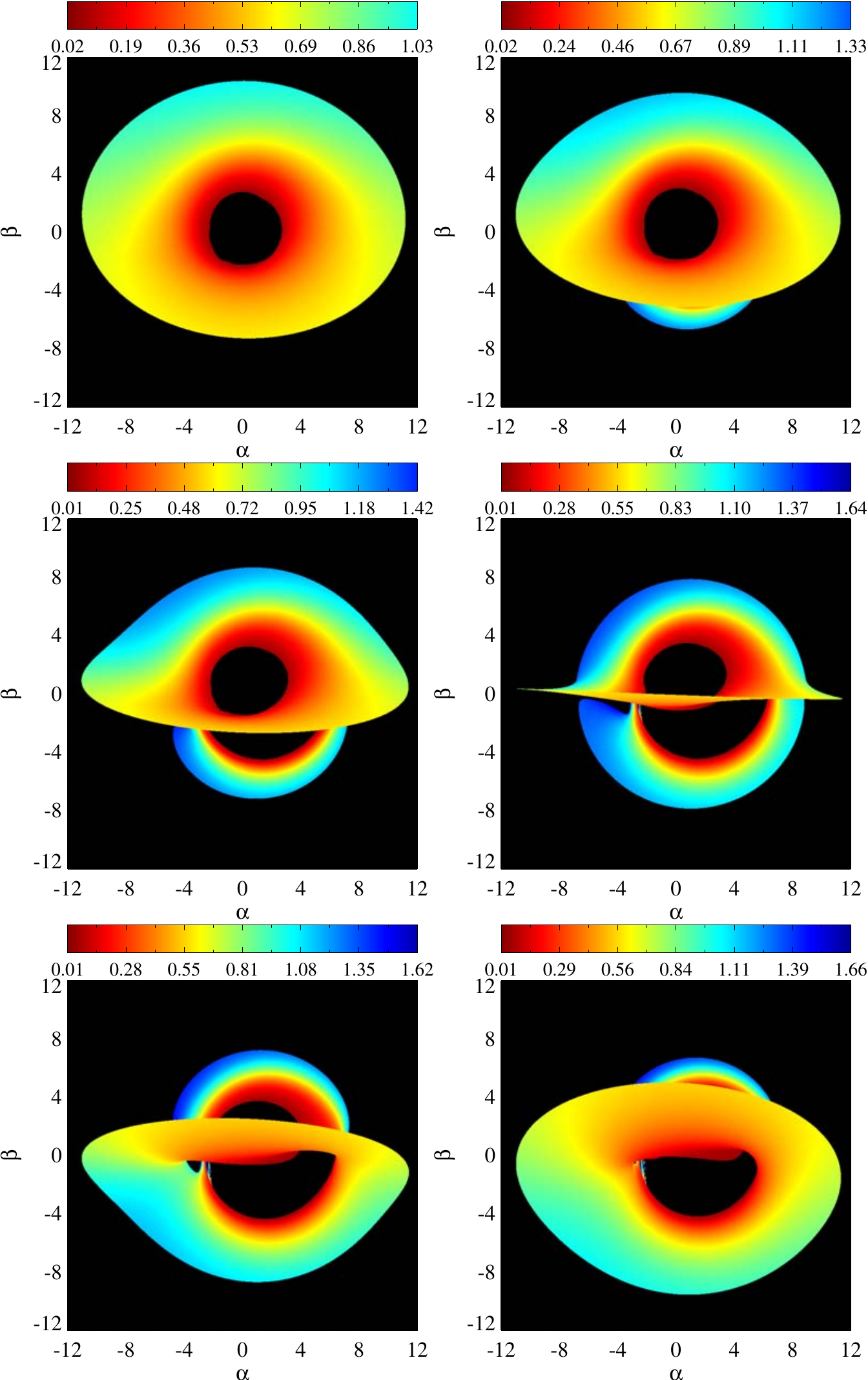}
\caption{\label{disk1} The images of a skewed disk, whose inner and
outer radii are 1.5 $r_\mathrm{g}$ and 10 $r_\mathrm{g}$. The
parameters: $\beta=30^\circ$, $\gamma_0=135^\circ$, black hole spin
$a$=0.998, inclination angles $\theta_{\mathrm{obs}}$ are
$15^\circ$, $30^\circ$, $45^\circ$, $60^\circ$, $75^\circ$ and
$90^\circ$ for panels a-f.}
\end{figure}

In Figure \ref{disk1}, we plot the images of a skewed accretion disk
that is composed by test particles falling freely into a black hole
along geodesics viewed from different inclinations. The initial
tiled angle of the disk is $\beta=30^\circ$. Due to the frame drag
effect, the particles drift into the black hole along spiral
orbits, instead of a straight lines. The orbits make a gradual
transition into the equatorial plane. The disk is significantly
warped as moving inward. In the figure, the false color represent
the redshift of emission coming from the disk surface. One can see
that the approaching and receding sides of the disk are no longer
the left and right sides, but the regions are farthest and nearest
with respect to the observer respectively.

\subsection{Stationary axisymmetric accretion flow}
\cite{tejeda2013} presented an analytic toy model to mimic the
stationary axisymmetric accretion flow of a
rotating cloud of non-interacting particles falling onto a Kerr black hole.
In which the streamlines are described analytically in terms of timelike
geodesics. Thus they solve the equations of motion with integral forms by
elliptic functions. However their results are completely different comparing
with ours. In addition, they just get the solutions for $r$ and $\mu$.

\begin{figure}
\centering
\includegraphics[width=0.4\textwidth,angle=0]{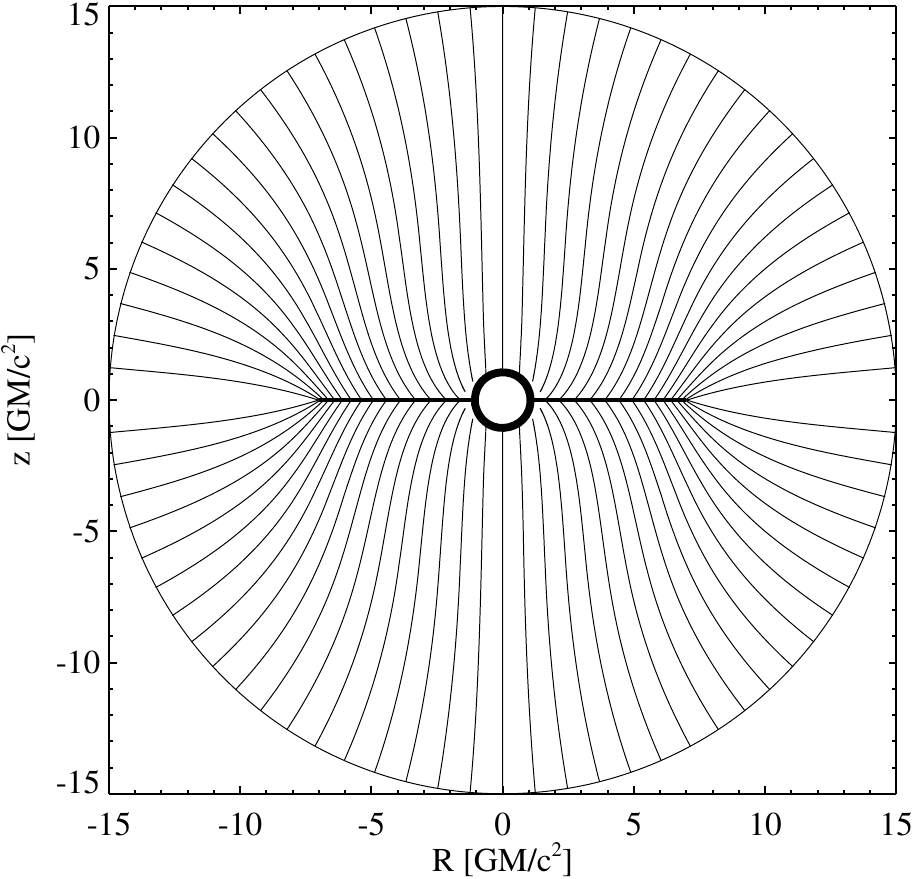}
\caption{\label{accretion} Streamlines of axisymmetric accretion flow. Parameters are:
black hole spin a=0.998; initial four velocities $u^r=0.35, u^\theta=0, u^\phi=-0.025$;
radius of the shell $r_0=15$ $r_\rms{g}$; $R=\sqrt{r^2+a^2}\sin\theta$, $z=r\cos\theta$.
The black circle and line represent the event horizon and accretion disk respectively.
Compare to Figure 1. of \cite{tejeda2013}.}
\end{figure}

As a check of the validation of our code, we use it to mimic the similar accretion
flow. Since the flow is axisymmetric, we just need to consider
the spacial projections onto the $r-\theta$ plane. The boundary of the flow
is a spherical shell at $r = r_0$ from which test particles are continuously injected.
On the shell, the four velocities of particles are taken to be constants, i.e.,
\begin{eqnarray}
  u^{r}(r_0,\theta)=
  u^{\theta}(r_0,\theta)=
  u^{\phi}(r_0,\theta)=const.,
\end{eqnarray}
with which and identity $g_{\mu\nu} u^\mu u^\nu=-1$, $u^t$ can be obtained.
$u^{\mu}$ at $r_0$ are taken as the initial conditions for the calculation of geodesics.
Using Equation (\ref{v_vs_dot_v}) the physical velocities
$\upsilon_r, \upsilon_\theta, \upsilon_\phi$ can be computed from $u^\mu$.
Then the constants of motion are uniquely determined. The results are
illustrated in Figure \ref{accretion}, which agree well with those of \cite{tejeda2013}.

\subsection{The tidal disruption of a ball}
As the final application of our code we use it to mimic a tidal disruption
event of a ball, which falls freely to the central black hole. To
use the code, we have to assume that the ball is consists of a set
of equal mass test particles without any interactions. At the
initial point the ball has a kick velocity, the physical components
of which measured under the LNRF reference are $\upsilon_r,
\upsilon_\theta, \upsilon_\phi$. And all of the particles share the
same initial velocities but different positions. Then with the given
initial conditions, each particle falls inward freely along a
geodesic trajectory.

\begin{figure}
\centering
\includegraphics[width=0.42\textwidth,angle=0]{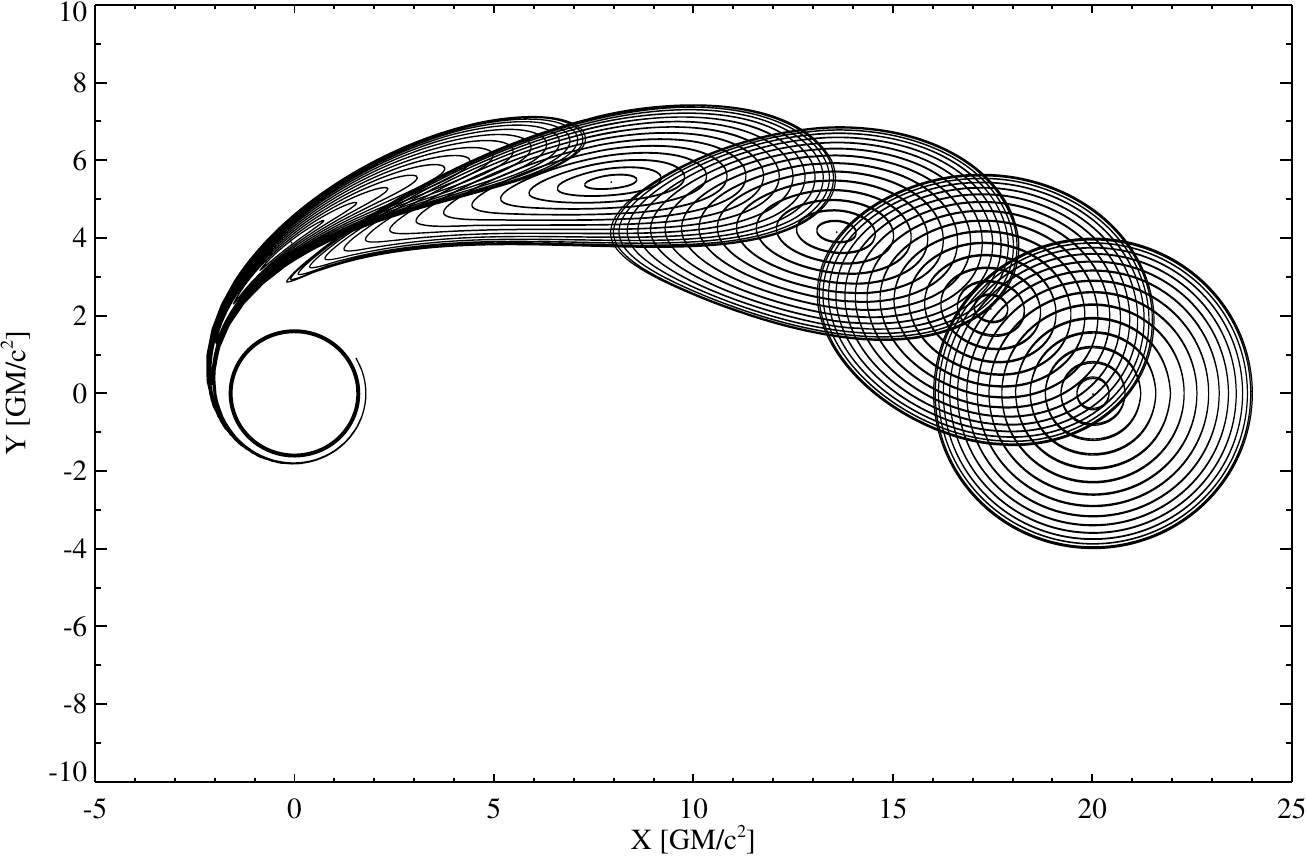}
\caption{\label{tidal_disruption} An assumed ball composed by a set
of massive test particles without interactions is disrupted by the
strong tidal force of a Kerr black hole. The black hole spin
$a=0.8$. The radius of the ball is 4 $r_\mathrm{g}$. The initial
coordinates are $r=20$ $r_\mathrm{g}$, $\theta=90^\circ$,
$\phi=0^\circ$. The coordinate times are $t=$ 0, 22.5, 45, 67.5, 90
for images form right to left. A circle in the left represents the
event horizon.}
\end{figure}

In Figure \ref{tidal_disruption}, we show the deformed images of the
ball for five different coordinate times. At initial moment, we
assume that the shape of ball is a regular sphere and the center of
the ball is located in the equatorial plane of the black hole. The
velocities are $\upsilon_r=-0.1, \upsilon_\theta=0,
\upsilon_\phi=0.1$. One can see that the shape of the ball is
significantly deformed and stretched as it approaches to the central
black hole due to the strong tidal disruption force. And the former
part is stretched more seriously than the latter part. The debris of
ball orbits around the black hole along a spiral trajectory and goes
inward slowly instead of falling into the black hole directly for
the frame drag effect. However, the picture illustrated by this
example is a toy model definitely.

\section{Discussion and conclusion}
\label{discconc}
We have developed a new fast public code named ynogkm for calculating
time-like geodesics under a K-N
spacetime, which is a direct extension of \cite{yangwang2012}. In ynogkm,
we adopt the same strategies used in ynogk, i.e., expressing all
coordinates and proper times as functions of a parameter $p$ and
calculating all elliptic integrals by Carlson's approach. The former
guarantees the convenience of the code in practice application and
the latter guarantees the fast speed of the code respectively. The
extension is involved in many more complicated cases.

In the expressions, we also use the Weierstrass' elliptic function
$\wp(z;g_2,g_3)$ and integral as investigated by many authors in the
literature. By this way they not only investigate the geodesic
motion itself but also the properties of the spacetime. While what
we discussed in this paper focus on the potential real applications
of the calculation of geodesic orbits in astrophysics. In order to
avoid the complex integrals, we also adopt the Jacobi's elliptic
functions $\mathrm{sn}(z|k^2),\mathrm{cn}(z|k^2)$ when equation
$R(r)=0$ has no real roots.

Since ynogkm uses the same strategies with ynogk, their speed are
almost the same, we do not present the speed test results. As discussed in
\cite{chan2013}, a powerful approach improving the speed of tracing
the trajectories of billions of photons in a curved spacetime is
based on the massively parallel algorithm and GPU graphic cards.
Their results show that this approach is two orders of magnitude
faster than the CPU-based tracing codes. Therefore the extension of
ynogkm from a serial program to a parallel program is the future
work.

To demonstrate the utility of ynogkm, we just apply it to six toy problems
and present the results simply. Its application to more complicated and
practical cases will be given in the future works.

\section*{Acknowledgments}
We acknowledge the anonymous referee for his/her valuable comments and
advices, which significantly improve the manuscript.
We acknowledge the financial supports from the National Natural
Science Foundation of China 11133006, 11163006, 11173054, the
National Basic Research Program of China (973 Program 2009CB824800),
and the Policy Research Program of Chinese Academy of Sciences
(KJCX2-YW-T24).

\bibliographystyle{aa}

\bibliography{ref}

\appendix
\section{The transformation matrix}
\label{matrixT} 
\begin{figure}
\begin{center}
\includegraphics[width=0.4\textwidth,angle=0]{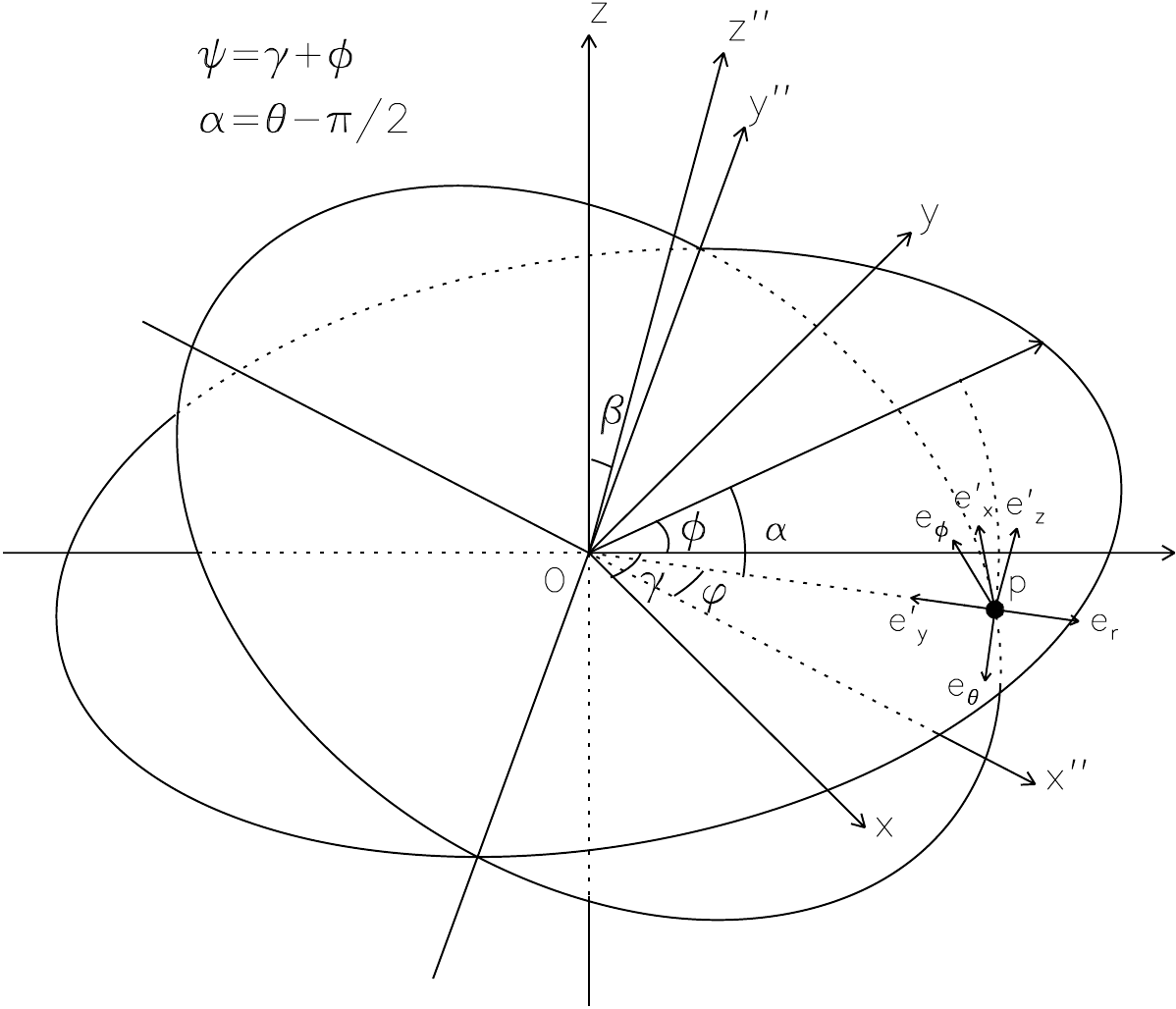}
\caption{\label{transformation} The geometry between the skewed disk
plane and the equatorial plane of a black hole. The boundary of the
disk is a ring, at which equal mass test particles are injected
continuously. At the initial position, indicated by p, there are two
orthonormal tetrads, i.e., \{p, $\mathbf{e}_r, \mathbf{e}_\theta,
\mathbf{e}_\phi$\} and \{p, $\mathbf{e}'_x, \mathbf{e}'_y,
\mathbf{e}'_z$\}. In order to compute the constants of motion, we
need to transform the initial physical velocities of a particle
$\upsilon_x', \upsilon_y',\upsilon_z'$, which are specified by \{p,
$\mathbf{e}'_x, \mathbf{e}'_y, \mathbf{e}'_z$\}, into \{p,
$\mathbf{e}_r, \mathbf{e}_\theta, \mathbf{e}_\phi$\}.}
\end{center}
\end{figure}

Here we discuss how to get the explicit expression
for the matrix $T$, which transform the physical velocities
$\upsilon_x', \upsilon_y', \upsilon_z'$ of a particle specified in
the reference of the tetrad $\mathbf{e}_x', \mathbf{e}_y',
\mathbf{e}_z'$ into the LNRF reference whose origin is fixed at the
same point. As shown in Figure \ref{transformation}, we have four
references, i.e., $R'$: \{p, $\mathbf{e}_x', \mathbf{e}_y',
\mathbf{e}_z'$\}, $R''$: \{$O, x'', y'', z''$\}, $R$: \{$O, x, y,
z$\}, and $R_{r\theta\phi}$: \{p, $\mathbf{e}_r, \mathbf{e}_\theta,
\mathbf{e}_\phi$\}.

The matrix of transformation $T_1$ from $R'$ into $R''$ can be
gotten directly,
\begin{eqnarray}
\label{T1}\left(
\begin{array}{c}
    \upsilon_x'' \\
    \upsilon_y'' \\
    \upsilon_z'' \\
\end{array}
\right)=\left(
\begin{array}{ccc}
    -\sin\varphi & -\cos\varphi & 0\\
    \cos\varphi & -\sin\varphi & 0\\
    0           &  0           & 1
\end{array}
\right)\left(
\begin{array}{c}
    \upsilon_x' \\
    \upsilon_y' \\
    \upsilon_z' \\
\end{array}
\right).
\end{eqnarray}
The transformation $T_2$ from $R''$ into $R$ is given by,
\begin{eqnarray}
\label{T2}\left(
\begin{array}{c}
    \upsilon_x \\
    \upsilon_y \\
    \upsilon_z \\
\end{array}
\right)=\left(
\begin{array}{ccc}
    \cos\gamma\cos\beta & -\sin\gamma & \cos\gamma\sin\beta \\
    \sin\gamma\cos\beta & \cos\gamma & \sin\gamma\sin\beta \\
    -\sin\beta          &  0          & \cos\beta
\end{array}
\right)\left(
\begin{array}{c}
    \upsilon_x'' \\
    \upsilon_y'' \\
    \upsilon_z'' \\
\end{array}
\right).
\end{eqnarray}
The transformation $T_3$ from $R$ into $R_{r\theta\phi}$ is given
by,
\begin{eqnarray}
\label{T3}\left(
\begin{array}{c}
    \upsilon_r \\
    \upsilon_\theta \\
    \upsilon_\phi \\
\end{array}
\right)=\left(
\begin{array}{ccc}
    \cos\psi\sin\theta & \sin\psi\sin\theta & \cos\theta  \\
    \cos\psi\cos\theta & \sin\psi\cos\theta & -\sin\theta \\
    -\sin\psi          &  \cos\psi          &  0
\end{array}
\right)\left(
\begin{array}{c}
    \upsilon_x \\
    \upsilon_y \\
    \upsilon_z \\
\end{array}
\right).
\end{eqnarray}

Thus the transformation $T$ from $R'$ into $R_{r\theta\phi}$ is
given by
\begin{eqnarray}
T = T_3 T_2 T_1,
\end{eqnarray}
and noting that $\psi=\gamma+\phi$, one has
\begin{eqnarray*}
T_3 T_2=\left(
\begin{array}{cc}
\sin\theta\cos\phi\cos\beta
-\cos\theta\sin\beta &  \sin\theta\sin\phi \\
 \cos\theta\cos\phi\cos\beta+\sin\theta\sin\beta   &   \cos\theta\sin\phi \\
-\sin\phi\cos\beta & \cos\phi
\end{array}
\right.\\
\left.
\begin{array}{c}
 \sin\theta\cos\phi\sin\beta+\cos\theta\cos\beta\\
 \cos\theta\cos\phi\sin\beta-\sin\theta\cos\beta \\
 -\sin\phi\sin\beta
\end{array}
\right).
\end{eqnarray*}
Finally one gets
\begin{eqnarray}
T =\left(
\begin{array}{ccc}
0     & -1      &   0\\
 -\sin\phi\sin\beta   &   0 & -\sqrt{1-\sin^2\phi\sin^2\beta} \\
\sqrt{1-\sin^2\phi\sin^2\beta} & 0 & -\sin\phi\sin\beta
\end{array}
\right).
\end{eqnarray}
In the reduction, the following identities are used
\begin{eqnarray}
\sin\theta = \frac{\cos\beta}{\sqrt{1-\sin^2\phi\sin^2\beta}},\\
\cos\theta = -\frac{\cos\phi\sin\beta}{\sqrt{1-\sin^2\phi\sin^2\beta}},\\
\sin\varphi = \frac{\sin\phi\cos\beta}{\sqrt{1-\sin^2\phi\sin^2\beta}},\\
\cos\varphi = \frac{\cos\phi}{\sqrt{1-\sin^2\phi\sin^2\beta}}.
\end{eqnarray}

\section{Taking $t$ and $\sigma$ to be the independent variable}
In some practical applications, one prefers using $t$ or $\sigma$ as
the independent variable than parameter $p$. Since we have expressed
all B-L coordinates and proper times as functions of the parameter
$p$, when a value of $t$ or $\sigma$ is given, there is an unique
$p$ corresponds to it. Namely, both the equations $t(p)=t_0$ and
$\sigma(p)=\sigma_0$ have one and only one real root, which is
denoted by $p_0$. Apparently, if we can solve these equations
efficiently and precisely to get $p_0$, we can take $t$ or $\sigma$
as the independent variable, for $p_0$ is obtained, the other three
coordinate $r, \mu$, and $\phi$ are also uniquely determined.

\begin{figure}
\begin{center}
\includegraphics[width=0.4\textwidth,angle=0]{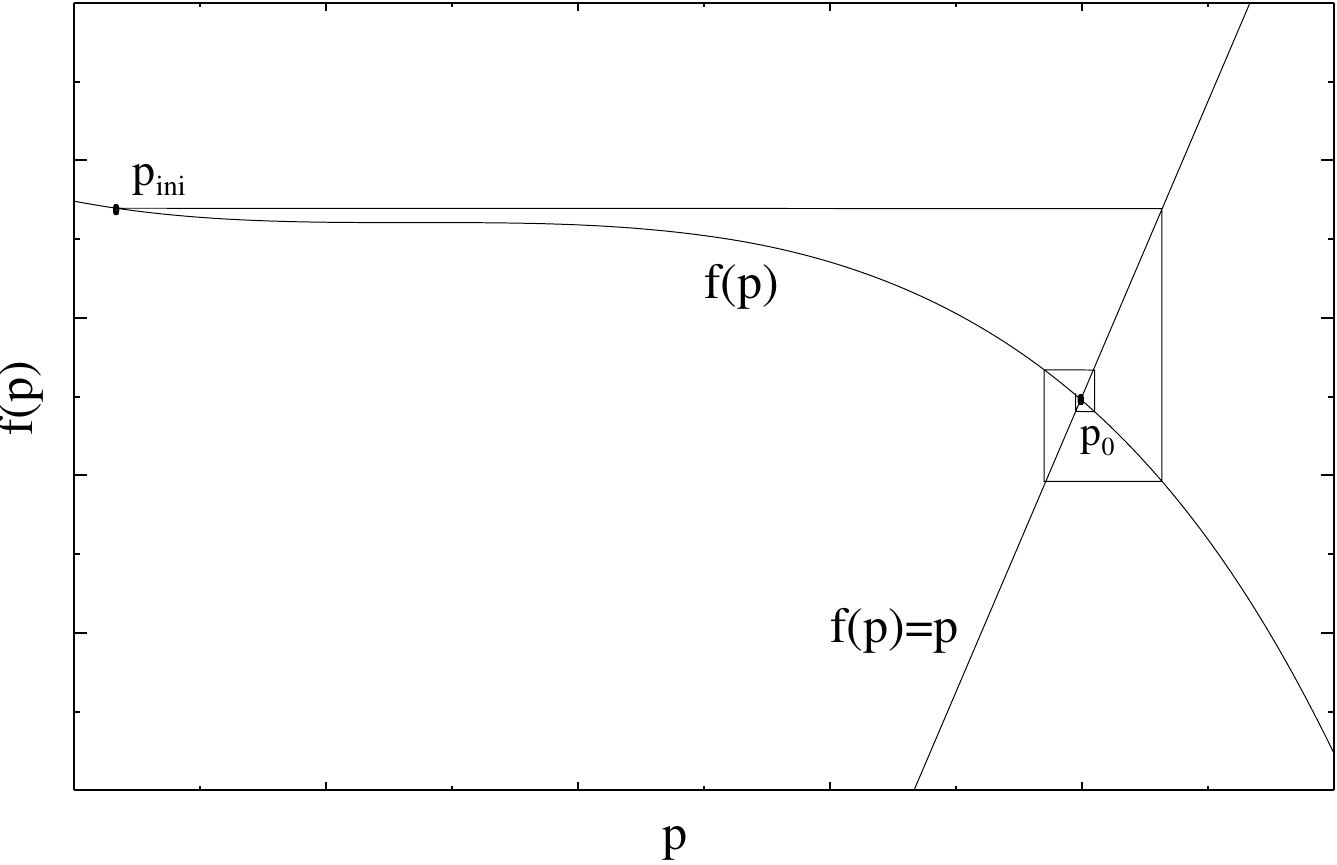}
\caption{\label{iterative} The curve of $f_\sigma(p)$ (or $f_t(p)$)
as function of $p$. $p_0$ is the unique real root of equation
$\sigma(p)=\sigma_0$ (or $t(p)=t_0$). Starting from an appropriate
initial value, $p_{\mathrm{ini}}$, one can approach to $p_0$ through
an iterative process.}
\end{center}
\end{figure}

Actually we can solve both the equations $t(p)=t_0$ and
$\sigma(p)=\sigma_0$ by bisection method or iterative method. From
the expressions for $t_r, t_\mu$ and $\sigma_r, \sigma_\mu$ given in
Section \ref{standartd_forms} we can rewrite functions $t(p)$ and
$\sigma(p)$ as
\begin{eqnarray}
   \sigma(p) &=& \overline{\sigma}(p)+C_\sigma p,\\
   t(p) &=& \overline{t}(p)+C_tp,
\end{eqnarray}
where the definitions of $C_\sigma$ and $C_t$ are given in table 7.
Thus for a given $\sigma_0$ and $t_0$, we have
\begin{eqnarray}
   p &=& \frac{\sigma_0-\overline{\sigma}(p)}{C_p} = f_\sigma(p),\\
   p &=& \frac{t_0-\overline{t}(p)}{C_t} = f_t(p).
\end{eqnarray}
We illustrate schematically how to solve these equations by
iterative method in Figure \ref{iterative}. To use the bisection
method, we define two new functions
\begin{eqnarray}
   F_\sigma(p) &=&  f_\sigma(p)-p,\\
   F_t(p) &=& f_t(p)-p.
\end{eqnarray}
From Figure \ref{iterative}, we can see that when $p<p_0$,
$F_\sigma(p)$ or $F_t(p)>$0; when $p>p_0$, $F_\sigma(p)$ or
$F_t(p)<$0. Thus through the use of bisection method, we can solve
the equations $F_\sigma(p)=0$ and $F_t(p) = 0$ immediately.

\begin{table}
\label{table7}
\begin{center}
    \begin{threeparttable}
\begin{tabular}{c|c}
  \MC{2}{c}{\textbf{Table} 7.} \\
  \hline \hline
    \ZZ{-5pt}{15pt} Case &  $C_\sigma,$ $C_{t}$  \\
   \hline
  \begin{minipage}[b]{1em}
             \begin{eqnarray*}
                  1\\
             \end{eqnarray*}
   \end{minipage}  &  \begin{minipage}[b]{22em} 
                         \begin{eqnarray*}
                             C_\sigma &=& a^2\mu_{\mathrm{tp}_1}^2+r^2_{\mathrm{tp}_1},\\
                             C_t &=& C_\sigma +  (2+e\varepsilon)(2+r_{\mathrm{tp}_1})-
                                    e^2\nonumber+A_{\mathrm{t}+}-A_{\mathrm{t}-},
                         \end{eqnarray*}
                   \end{minipage}
               \\ \hline
  \begin{minipage}[b]{1em}
             \begin{eqnarray*}
                  2,\;3
             \end{eqnarray*}
   \end{minipage}   &   \begin{minipage}[b][\height]{22em} 
                     \begin{eqnarray*}
                          C_\sigma &=& a^2\mu_{\mathrm{tp}_1}^2,\; C_t = C_\sigma +\frac{1}{\sqrt{|1-m^2|}}[2(2+e\varepsilon)-e^2],
                     \end{eqnarray*}
                 \end{minipage}
            \\
          \hline
   \begin{minipage}[b]{1em}
             \begin{eqnarray*}
               4,\;5
             \end{eqnarray*}
   \end{minipage}    &   \begin{minipage}[b][\height]{22em} 
                         \begin{eqnarray*}
                             C_\sigma &=& b_1^2+\frac{a^2\mu^2_0}{2}\tnote{1},\;\;\; C_t = C_\sigma + (2+e\varepsilon)(2-b_1/b_0)-e^2.
                         \end{eqnarray*}
                  \end{minipage}
           \\
  \hline\hline
\end{tabular}
\begin{tablenotes}
  \footnotesize
  \item[1]  $\mu_0 = \sqrt{q/(q+\lambda^2)}$.
\end{tablenotes}
\end{threeparttable}
\end{center}
\end{table}


\end{document}